\documentclass[longauth, noline]{aa}
\usepackage{graphicx}
\usepackage{txfonts,textcomp}
\usepackage{booktabs}
\usepackage{natbib}
\usepackage{bm}
\usepackage{orcidlink}
\begin{document}

    \authorrunning{Vince et al.}
    
    \titlerunning{Multiband optical variability of Ton 599}

    \title{Multiband optical variability on diverse timescales of the blazar Ton 599 from 2011 to 2023}

    \author{
        O. Vince\inst{\ref{1} \and \ref{2}}\orcidlink{0009-0008-5761-3701},
        C. M. Raiteri\inst{\ref{3}}\orcidlink{0000-0003-1784-2784},
        M. Villata\inst{\ref{3}}\orcidlink{0000-0003-1743-6946},
        A. C. Gupta\inst{\ref{4} \and \ref{5}}\orcidlink{0000-0002-9331-4388},
        J. Kova\v cevi\'c-Doj\v cinovi\'c\inst{\ref{1}}\orcidlink{0000-0002-1593-707X},
        M. Laki\'cevi\'c\inst{\ref{1}}\orcidlink{0000-0002-8231-0963},
        L. \v C. Popovi\'c\inst{\ref{1} \and \ref{6}}\orcidlink{0000-0003-2398-7664},
        P. Kushwaha\inst{\ref{7}}\orcidlink{0000-0001-6890-2236},
        D. O. Mirzaqulov\inst{\ref{8}}\orcidlink{0000-0003-0570-6531},
        S. A. Ehgamberdiev\inst{\ref{8} \and \ref{9}}\orcidlink{0000-0001-9730-3769},
        D. Carosati\inst{\ref{10} \and \ref{11}},
        S. G. Jorstad\inst{\ref{12} \and \ref{13}}\orcidlink{0000-0001-6158-1708},
        A. P. Marscher\inst{\ref{12}}\orcidlink{0000-0001-7396-3332},
        Z. R. Weaver\inst{\ref{12}}\orcidlink{0000-0001-6314-0690},
        J. R. Webb\inst{\ref{14}}\orcidlink{0000-0001-6078-2022},
        P. S. Smith\inst{\ref{15}}\orcidlink{0000-0002-5083-3663},
        W. Chen\inst{\ref{16}}\orcidlink{0000-0003-0262-272X},
        A. Tsai\inst{\ref{16} \and \ref{16a}}\orcidlink{0000-0002-3211-4219},
        H. Lin\inst{\ref{16}},
        G. A. Borman\inst{\ref{17}}\orcidlink{0000-0002-7262-6710},
        T. S. Grishina\inst{\ref{13}}\orcidlink{0000-0002-3953-6676},
        V. A. Hagen-Thorn\inst{\ref{13}}\orcidlink{0000-0002-6431-8590},
        E. N. Kopatskaya\inst{\ref{13}}\orcidlink{0000-0001-9518-337X},
        E. G. Larionova\inst{\ref{13}}\orcidlink{0000-0002-2471-6500},
        V. M. Larionov\inst{\ref{13}},
        L. V. Larionova\inst{\ref{13}}\orcidlink{0000-0002-0274-1481},
        D. A. Morozova\inst{\ref{13}}\orcidlink{0000-0002-9407-7804},
        S. S. Savchenko\inst{\ref{13} \and \ref{13a}}\orcidlink{0000-0003-4147-3851},
        I. S. Troitskiy\inst{\ref{13}}\orcidlink{0000-0002-4218-0148},
        Y. V. Troitskaya\inst{\ref{13}}\orcidlink{0000-0002-9907-9876},
        A. A. Vasilyev\inst{\ref{13}}\orcidlink{0000-0002-8293-0214},
        A. V. Zhovtan\inst{\ref{17}},
        E. V. Shishkina\inst{\ref{13}}\orcidlink{0009-0002-2440-2947},
        O. M. Kurtanidze\inst{\ref{18} \and \ref{19} \and \ref{20} \and \ref{21} \and \ref{22}}\orcidlink{0000-0001-5385-0576},
        M. G. Nikolashvili\inst{\ref{18} \and \ref{21} \and \ref{22}}\orcidlink{0000-0003-0408-7177},
        S. O. Kurtanidze\inst{\ref{18} \and \ref{22}}\orcidlink{0000-0002-0319-5873},
        R. Ivanidze\inst{\ref{18}}\orcidlink{0009-0005-7297-8985},
        J. A. Acosta-Pulido\inst{\ref{23} \and \ref{24}}\orcidlink{0000-0002-0433-9656},
        M. I. Carnerero\inst{\ref{3}}\orcidlink{0000-0001-5843-5515},
        G. Damljanovi\'c\inst{\ref{1}}\orcidlink{0000-0002-6710-6868},
        M. Stojanovi\'c\inst{\ref{1}}\orcidlink{0000-0002-4105-7113},
        M. D. Jovanovi\'c\inst{\ref{1}}\orcidlink{0000-0003-4298-3247},
        V. V. Vlasyuk\inst{\ref{25}}\orcidlink{0009-0002-6596-7274},
        O. I. Spiridonova\inst{\ref{25}}\orcidlink{0009-0007-7315-3090},
        A. S. Moskvitin\inst{\ref{25}},
        T. Pursimo\inst{\ref{26}}\orcidlink{0000-0002-5578-9219},
        D. Elsässer\inst{\ref{27} \and \ref{28}}\orcidlink{0000-0001-6796-3205},
        M. Feige\inst{\ref{27}},
        L. Kunkel\inst{\ref{27}},
        J. Ledermann\inst{\ref{27}},
        %K. Mannheim\inst{\ref{27} \and \ref{29}}\orcidlink{0000-0002-2950-6641},
        D. Reinhart\inst{\ref{27}},
        A. Scherbantin\inst{\ref{27}},
        K. Schoch\inst{\ref{27}},
        R. Steineke\inst{\ref{27}},
        C. Lorey\inst{\ref{27}}\orcidlink{0009-0002-5220-2993},
        I. Agudo\inst{\ref{30}}\orcidlink{0000-0002-3777-6182},
        J. E. Pedrosa\inst{\ref{31} \and \ref{30}}\orcidlink{0000-0002-4131-655X},
        F. J. Aceituno\inst{\ref{30}}\orcidlink{0000-0001-8074-4760},
        G. Bonnoli\inst{\ref{32}}\orcidlink{0000-0003-2464-9077},
        V. Casanova\inst{\ref{30}}\orcidlink{0000-0003-2036-8999},
        D. Morcuende\inst{\ref{30}}\orcidlink{0000-0001-9400-0922},
        A. Sota\inst{\ref{30}}\orcidlink{0000-0002-9404-6952},
        V. Bozhilov\inst{\ref{33}}\orcidlink{0000-0002-3117-7197},
        A. Valcheva\inst{\ref{33}}\orcidlink{0000-0002-2827-4105},
        E. Zaharieva\inst{\ref{33}}\orcidlink{0000-0001-7663-4489},
        M. Minev\inst{\ref{34}}\orcidlink{0000-0002-5702-5095},
        A. Strigachev\inst{\ref{34}},
        R. Bachev\inst{\ref{34}}\orcidlink{0000-0002-0766-864X},
        B. Mihov\inst{\ref{34}},
        L. Slavcheva-Mihova\inst{\ref{34}},
        A. C. Sadun\inst{\ref{35}}\orcidlink{0000-0001-8086-7242},
        A. Takey\inst{\ref{36}}\orcidlink{0000-0003-1423-5516},
        A. Shokry\inst{\ref{36}}\orcidlink{0000-0003-3052-6145},
        M. A. El-Sadek\inst{\ref{36}},
        A. Marchini\inst{\ref{37}}\orcidlink{0000-0003-3779-6762},
        G. Verna\inst{\ref{37}}\orcidlink{0000-0001-5916-9028}
    }
    \institute{
        Astronomical Observatory, Volgina 7, 11060 Belgrade, Serbia, \email{ovince@aob.rs}\label{1}
        \and
        Shanghai Astronomical Observatory, Chinese Academy of Sciences, 80 Nandan Road, Shanghai 200030, People's Republic of China\label{2}
        \and
        INAF, Osservatorio Astrofisico di Torino, via Osservatorio 20, I-10025 Pino Torinese, Italy\label{3}
        \and
        Aryabhatta Research Institute of Observational Sciences (ARIES), Manora Peak, Nainital 263001, India\label{4}
        \and
        Xinjiang Astronomical Observatory, CAS, 150 Science-1 Street, Urumqi 830011, China\label{5}
        \and
        Department of Astronomy, University of Belgrade - Faculty of Mathematics, Studentski trg 16, 11000 Belgrade, Serbia\label{6}
        \and
        Department of Physical Sciences, Indian Institute of Science Education and Research (IISER) Mohali, Knowledge City, Sector 81, SAS Nagar, Manauli 140306, India\label{7}
        \and
        Ulugh Beg Astronomical Institute, Astronomy Street 33, Tashkent 100052, Uzbekistan\label{8}
        \and
        National University of Uzbekistan, Tashkent 100174, Uzbekistan\label{9}
        \and
        EPT Observatories, Tijarafe, La Palma, Spain\label{10}
        \and
        INAF, TNG Fundaci\'on Galileo Galilei, La Palma, Spain\label{11}
        \and
        Institute for Astrophysical Research, Boston University, 725 Commonwealth Avenue, Boston, MA 02215, USA\label{12}
        \and
        Saint Petersburg State University, 7/9 Universitetskaya nab., St. Petersburg, 199034 Russia\label{13}
        \and
        Pulkovo Observatory, St.Petersburg, 196140, Russia\label{13a}
        \and
        Department of Physics Florida International University and the SARA Observatory, Miami, FL 33199, USA\label{14}
        \and
        Steward Observatory, University of Arizona, 933 N. Cherry Ave., Tucson, AZ 85721 USA\label{15}
        \and
        Institute of Astronomy, National Central University, 300 Zhongda Rd., Taoyuan 320317, Taiwan\label{16}
        \and
        Taiwan Astronomical Research Alliance, 300 Zhongda Rd., Taoyuan 320317, Taiwan\label{16a}
        \and
        Crimean Astrophysical Observatory RAS, P/O Nauchny, 298409, Russia\label{17}
        \and
        Abastumani Observatory, Mt. Kanobili, 0301 Abastumani, Georgia\label{18}
        \and
        Max-Planck-Institut f\"ur Radioastronomie, Auf dem H\"ugel 69, 53121 Bonn, Germany\label{19}
        \and
        Engelhardt Astronomical Observatory, Kazan Federal University, Tatarstan, Russia\label{20}
        \and
        Center for Astrophysics, Guangzhou University, Guangzhou, 510006, China\label{21}
        \and
        Landessternwarte, Zentrum f\"ur Astronomie der Universitat Heidelberg, Knigstuhl 12, 69117, Heidelberg, Germany\label{22}
        \and
        Instituto de Astrofisica de Canarias (IAC), E-38200 La Laguna, Tenerife, Spain\label{23}
        \and
        Universidad de La Laguna (ULL), Departamento de Astrofisica, E-38206 La Laguna, Tenerife, Spain\label{24}
        \and
        Special Astrophysical Observatory of Russian Academy of Sciences, Nyzhnij Arkhyz, Karachai-Circassia, Russia, 369167\label{25}
        \and
        Nordic Optical Telescope, Apartado 474 E-38700 Santa Cruz de La Palma, Santa Cruz de Tenerife Spain\label{26}
        \and
        Hans-Haffner-Sternwarte (Hettstadt), Naturwissenschaftliches Labor f\"ur Sch\"uler am FKG; Friedrich-Koenig-Gymnasium, D-97082 W\"urzburg, Germany\label{27}
        \and
        Astroteilchenphysik, TU Dortmund, Otto-Hahn-Str. 4A, D-44227 Dortmund, Germany\label{28}
        \and
        %Lehrstuhl f\"ur Astronomie, Universitat W\"urzburg, D-97074 W\"urzburg, Germany\label{29}
        %\and
        Instituto de Astrof\'isica de Andaluc\'ia, IAA-CSIC, Glorieta de la Astronom\'ia s/n, E-18008 Granada, Spain\label{30}
        \and
        Center for Astrophysics - Harvard \& Smithsonian, 60 Garden Street, Cambridge, MA 02138 USA\label{31}
        \and
        INAF Osservatorio Astronomico di Brera, Via E. Bianchi 46, 23807 Merate (LC), Italy\label{32}
        \and
        Department of Astronomy, Faculty of Physics, Sofia University `St. Kliment Ohridski' 5 James Bourchier Blvd., BG-1164 Sofia, Bulgaria\label{33}
        \and
        Institute of Astronomy and National Astronomical Observatory, Bulgarian Academy of Sciences, 72 Tsarigradsko shosse Blvd., 1784 Sofia, Bulgaria\label{34}
        \and
        Department of Physics, University of Colorado Denver, Denver, Colorado 80204, USA\label{35}
        \and
        National Research Institute of Astronomy and Geophysics (NRIAG), 11421 Helwan, Cairo, Egypt\label{36}
        \and
        Astronomical Observatory, University of Siena, Via Roma 56, 53100, Siena, Italy\label{37}
    }

    \date{Received August 30, 2025; accepted September 01, 2025}

    \abstract
    {We analyze the optical variability of the flat-spectrum radio quasar (FSRQ) Ton 599 using BVRI photometry from the Whole Earth Blazar Telescope (WEBT) collaboration (2011–2023), complemented by photometric and spectroscopic data from the Steward Observatory monitoring program.}
    {We aim to characterize short- and long-term optical variability — including flux distributions, intranight changes, color evolution, and spectra — to constrain physical parameters and processes in the central engine of this active galactic nucleus (AGN).}
    {We tested flux distributions in each filter against normal and log-normal models and explored the root mean square (RMS)–flux relation. We derived power spectral densities (PSDs) to assess red-noise behavior. We quantified intranight variability using a $\chi^2$ test and fractional variability.
    From variability timescales, we estimated the emitting region size and magnetic field. Long-term variability was studied by segmenting the light curve into 12 intervals and analyzing flux statistics. For multi-filter flares, we computed spectral slopes, redshift-corrected fluxes, and monochromatic luminosities. Color-magnitude and color-time diagrams traced color evolution over different flux regimes and timescales. From low-flux spectra, we measured Mg II line properties (correcting for Fe II) to estimate the black hole mass via single-epoch scaling.}
    {During the monitoring period, Ton 599 showed strong optical variability. Log-normal distributions fit the fluxes better than normal ones, and all bands display a positive RMS–flux relation. The PSDs follow red-noise trends. Intranight variability is detected, with derived timescales constraining the emission region and magnetic field. The R band reaches a peak flux of 23.5 mJy, corresponding to a monochromatic luminosity of $\log(\nu L_{\nu}) = 48.48 \;[\mathrm{erg\,s^{-1}}]$. Color-magnitude diagrams reveal a redder-when-brighter trend at low fluxes (thermal dominance), achromatic behavior at intermediate levels (possibly due to jet orientation changes), and a bluer-when-brighter trend at high fluxes (synchrotron dominance). While long-term color changes are modest, short-term variations are significant, with a negative correlation between the amplitude of color changes and the average flux. The estimated supermassive black hole mass is on the order of $10^8 M_\odot$, which is in agreement with previous estimates.
    }
    {Our results underscore the complexity of blazar variability, pointing to multiple emission processes at work. The joint photometric and spectroscopic approach constrains key physical parameters and deepens our understanding of the blazar central engine.}
    
    \maketitle
	\nolinenumbers %% Added by PK for ArXiv

\section{Introduction}

\indent Blazars are a class of active galactic nuclei (AGNs) whose powerful relativistic jets are oriented very close to our line of sight \citep{1995PASP..107..803U}. This alignment causes significant Doppler boosting of the emitted radiation, leading to extreme observational properties such as high luminosity, rapid and large-amplitude variability across the entire electromagnetic spectrum, and strong, variable polarization \citep{1995PASP..107..803U, 1966ApJ...146..964K}. The variability of blazars occurs on a wide range of timescales, from minutes (intraday variability, IDV) to years (long-term variability, LTV), and is generally stochastic in nature, with larger amplitudes on longer timescales \citep[e.g.,][]{2008ApJ...689...79C, 2009ApJ...704.1689C, 2017MNRAS.464.2046K}.
\\
\indent Blazars are traditionally classified into two main categories based on their optical spectra: BL Lacertae objects (BL Lacs) that exhibit weak or no emission lines, and flat-spectrum radio quasars (FSRQs) that are characterized by strong, broad emission lines \citep{1978PhyS...17..265B, 1991ApJS...76..813S}. Their spectral energy distribution (SED) typically shows two broad humps \citep{1998MNRAS.299..433F}. The low-energy component, from radio to optical/UV frequencies, is dominated by synchrotron radiation from relativistic electrons in the jet \citep{1966ApJ...146..964K}. The high-energy component, from X-rays to $\gamma$-rays, is most commonly explained by inverse Compton (IC) scattering of lower-energy photons \citep{1985ApJ...298..114M, 1994ApJ...421..153S, 1992A&A...256L..27D, 1996A&AS..120C.503G}.
\\
\indent Based on the position of the synchrotron peak frequency of the SED, blazars are divided into three categories: low-synchrotron peaked (LSP) blazars ($\nu_{\mathrm{peak}} < 10^{14}$~Hz), intermediate-synchrotron peaked (ISP) blazars (${10^{14} < \nu_{\mathrm{peak}} < 10^{15}}$~Hz), and high-synchrotron peaked (HSP) blazars (${10^{15} < \nu_{\mathrm{peak}} < 10^{17}}$~Hz) \citep{2011ApJ...743..171A}. The FSRQs are mostly LSPs, while BL Lacs can be found in all three categories and are, accordingly, called low-, intermediate-, and high-frequency-peaked BL Lac objects \citep{2010ApJ...716...30A}. There are blazars for which the synchrotron peak is at $\nu_{\mathrm{peak}} > 10^{17}$~Hz; these are called extremely high-synchrotron-peaked BL Lacs \citep[EHBL;][]{2019MNRAS.486.1741F}.
\\
\indent The subject of this study, Ton~599 (also known as 1156+295), is a luminous FSRQ \citep{2007ApJS..171...61H} at a redshift $z = 0.72469$ \citep{2010MNRAS.405.2302H}. It is powered by a supermassive black hole with a mass of approximately $9 \times 10^8 M_{\odot}$ \citep[][]{2019PhDT.......127K}. VLBI observations of its parsec-scale jet have revealed highly superluminal motion, allowing for the determination of a Doppler factor of $\delta = 12 \pm 3$ and a jet viewing angle of $\theta \le 2.5^{\circ}$ \citep{2017ApJ...846...98J}. At $\gamma$-ray energies, the source was first detected  by the Compton Gamma-Ray Observatory (CGRO) and listed in the second Energetic Gamma Ray Experiment Telescope (EGRET) catalog \citep{1995ApJS..101..259T}. Ton 599 was confirmed as a gamma-ray source by the Fermi Large Area Telescope (LAT), a detection made possible by the instrument's improved sensitivity within the first three months of its mission \citep{2010ApJS..188..405A}. Ton~599 has a rich history of dramatic activity. Early observations documented rapid flux changes, such as a 0.05 mag brightening in the B band in just 29 minutes \citep{1984ApJ...277...77G}. A long-term study by \cite{2006PASJ...58..797F} found large variations of nearly 5 magnitudes and suggested periodicities of 1.58 and 3.55 years. More recently, a major multiwavelength outburst in late 2017 led to its first detection at very high energies by VERITAS and MAGIC \citep{2017ATel11061....1M, 2017ATel11075....1M}. Detailed studies of this event focused on constraining the jet's physical parameters and the location of the emission region, revealing a complex connection between the nonthermal continuum and the broad emission lines \citep{2022ApJ...926..180H, 2018ApJ...866..102P, 2019ApJ...871..101P}. The source's activity continued, with a record-breaking $\gamma$-ray flare in January 2023, which has also been a subject of detailed broadband spectral modeling \citep{2020MNRAS.492...72P, 2024MNRAS.529.1356M}. Other recent works have investigated its long-term radio kinematics and magnetic field properties \citep{2014MNRAS.445.1636R, 2021A&A...651A..74K}.
\\
\indent Given this history of complex and powerful outbursts, continuous and detailed long-term monitoring is essential to fully understand the physical processes at play. In this paper, we present one of the most extensive optical photometric studies of Ton~599 to date, utilizing a rich dataset collected from November 2011 to September 2023. The core of our analysis is based on BVRI photometry from the Whole Earth Blazar Telescope (WEBT) collaboration, which we complement with crucial long-term photometric and spectroscopic data from the Steward Observatory. Our primary goal is to perform a comprehensive characterization of the blazar’s multiband optical variability on diverse timescales in order to investigate the physical parameters and processes in the central part of this AGN. To this end, we test the flux distributions against normal and log-normal models and explore the root mean square (RMS)–flux relation. We quantify intranight changes using $\chi^2$ tests and fractional variability to estimate the emitting region size and magnetic field strength. Finally, we trace the source's complex color evolution using color–magnitude and color–time diagrams. Furthermore, we leverage the low-flux optical spectra to measure the properties of the Mg~II emission line, after correcting for Fe~II contamination, to provide an independent estimate of the central black hole mass via single-epoch scaling.
\\
\indent In Sect. 2, we describe the extensive optical photometric data collected from various observatories worldwide under the WEBT consortium. In Sect. 3, we present our results related to flux, color, and spectral studies. In Sect. 4, we discuss our findings and in Sect. 5 we summarize our conclusions.

\section{Optical data}

\indent For the analysis of the emission behavior of the blazar Ton 599, we used optical photometric data obtained within the framework of the Whole Earth Blazar Telescope (WEBT)\footnote{\url{https://www.oato.inaf.it/blazars/webt/}} project \citep[e.g.,][]{2002A&A...390..407V,2008A&A...481L..79V,2017Natur.552..374R,2020MNRAS.492.3829L,2022Natur.609..265J}. The analysis covers November 2011 to September 2023 (from JD 2455875.6141 to JD 2460192.30862). Photometric observations were made in $B$, $V$, $R$, and $I$ Johnson-Cousins filters. The total number of data points in the above filter bands included in the final processed light curves are 1489, 2089, 6625, and 1851, respectively. Table \ref{tab: observatories} lists the observatories, the telescope diameters that were used for the observations, and the total number of data points provided by individual instruments included in the analysis.
\\
\indent Instrumental magnitudes were reduced using master bias, dark (where appropriate) and flat-field calibration images. Cosmetic corrections included masking hot and dead pixels, removal of cosmic rays and fringing in the I filter where fringing is most pronounced. Photometric calibration was done with standard stars listed on the official WEBT website\footnote{\url{https://www.oato.inaf.it/blazars/webt/1156295-4c-29-45/}}.
\\
\indent The optical light curves were carefully processed to ensure homogeneity and precision across all four BVRI bands. The following steps were applied: (i) removal of outliers, including those within intranight sequences; (ii) binning of closely spaced and noisy measurements from the same dataset; (iii) inspection of data points with large uncertainties — when supported by independent observations taken on the same or nearby nights, the uncertainties were revised accordingly, otherwise the data points were discarded; (iv) imposition of a minimum photometric uncertainty of 0.01 mag and exclusion of all measurements with uncertainties larger than 0.15 mag; and (v) correction for systematic offsets between different datasets, applied when necessary to minimize inter-dataset discrepancies, while avoiding artificial time-dependent shifts.
\\
\indent Throughout the paper we use magnitudes corrected for Galactic extinction. Galactic extinction values were taken from the NASA/IPAC Extragalactic Database\footnote{\url{http://ned.ipac.caltech.edu}}, with values of 0.072, 0.054, 0.043, and 0.030 for the $B$, $V$, $R$, and $I$ bandpasses, respectively. For the conversion of magnitudes into fluxes, we use zero-magnitude flux densities provided by \citet{1998A&A...337..321B}. Throughout the paper, times are expressed as JD-2455000. Assuming a flat Universe ( $H_{0} = 68 \,\mathrm{km\,s^{-1}\,Mpc^{-1}}$, $\Omega_{m}\!=\!0.3$ and $\Omega_{\Lambda}\!=\!0.7$),  the luminosity distance at redshift $z = 0.72469$ is $D_{\mathrm{L}} \approx 4.6$ Gpc.

\section{Analysis and results}

\begin{figure}
\centering
\includegraphics[height=10cm,width=\columnwidth]{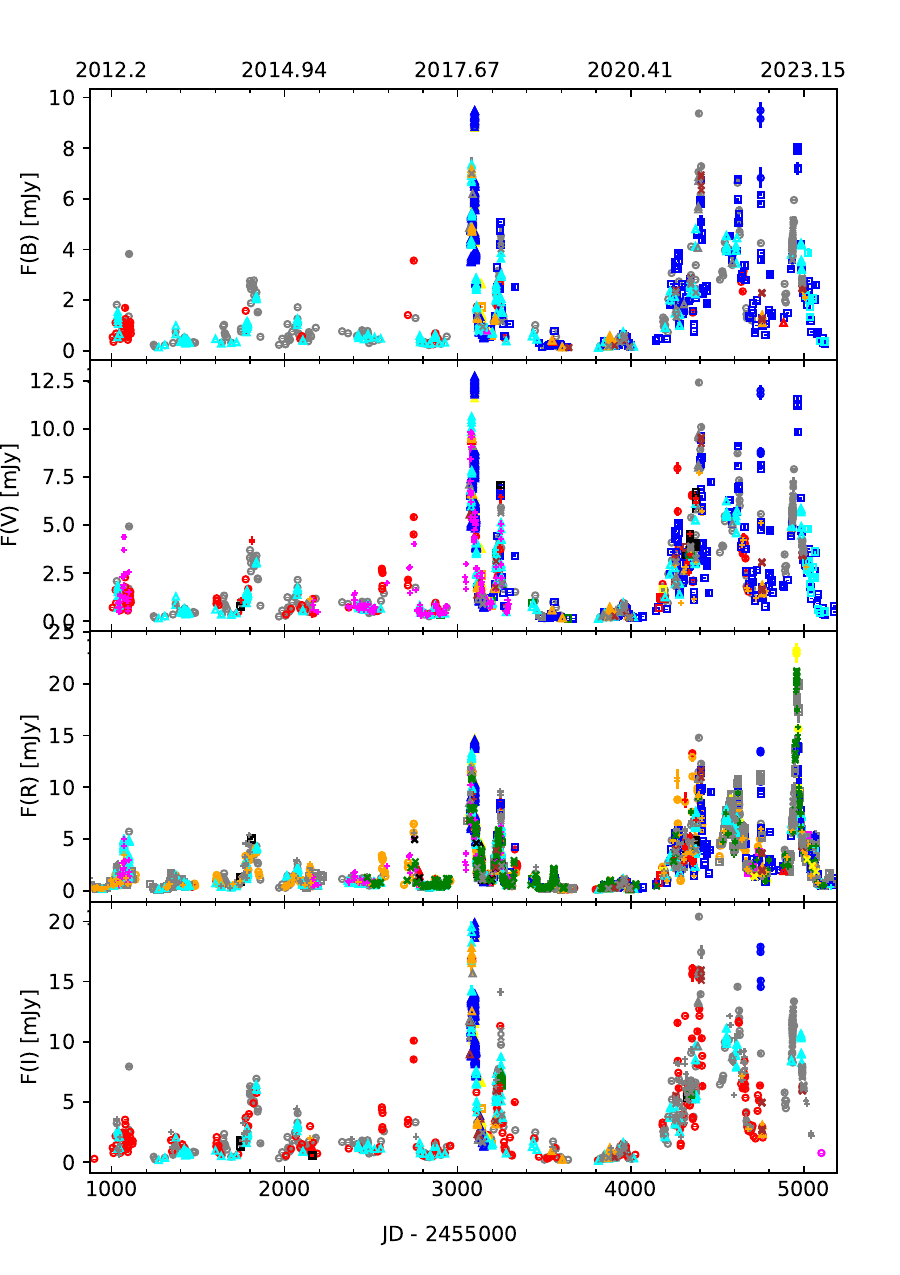}
\caption{Flux density evolution of the blazar Ton 599 from November 2011 to September 2023 in B, V, R, and I filters. Colors and symbols in the plots are explained in Table \ref{tab: observatories}.} 
\label{fig: F-PD-EVPA_vs_JD}
\end{figure}

\indent Figure \ref{fig: F-PD-EVPA_vs_JD}, viewed from top to bottom, shows the light curves of Ton 599 blazar in the B, V, R, and I filters, respectively, for the whole time period under consideration. Data points from different observatories and instruments are marked with different colors and symbols, as  described in Table \ref{tab: observatories}. 
\\

\subsection{Global flux properties} \label{subsec: Global_flux_properties}

\begin{figure}
\centering    
\includegraphics[width=0.5\textwidth,keepaspectratio=true]{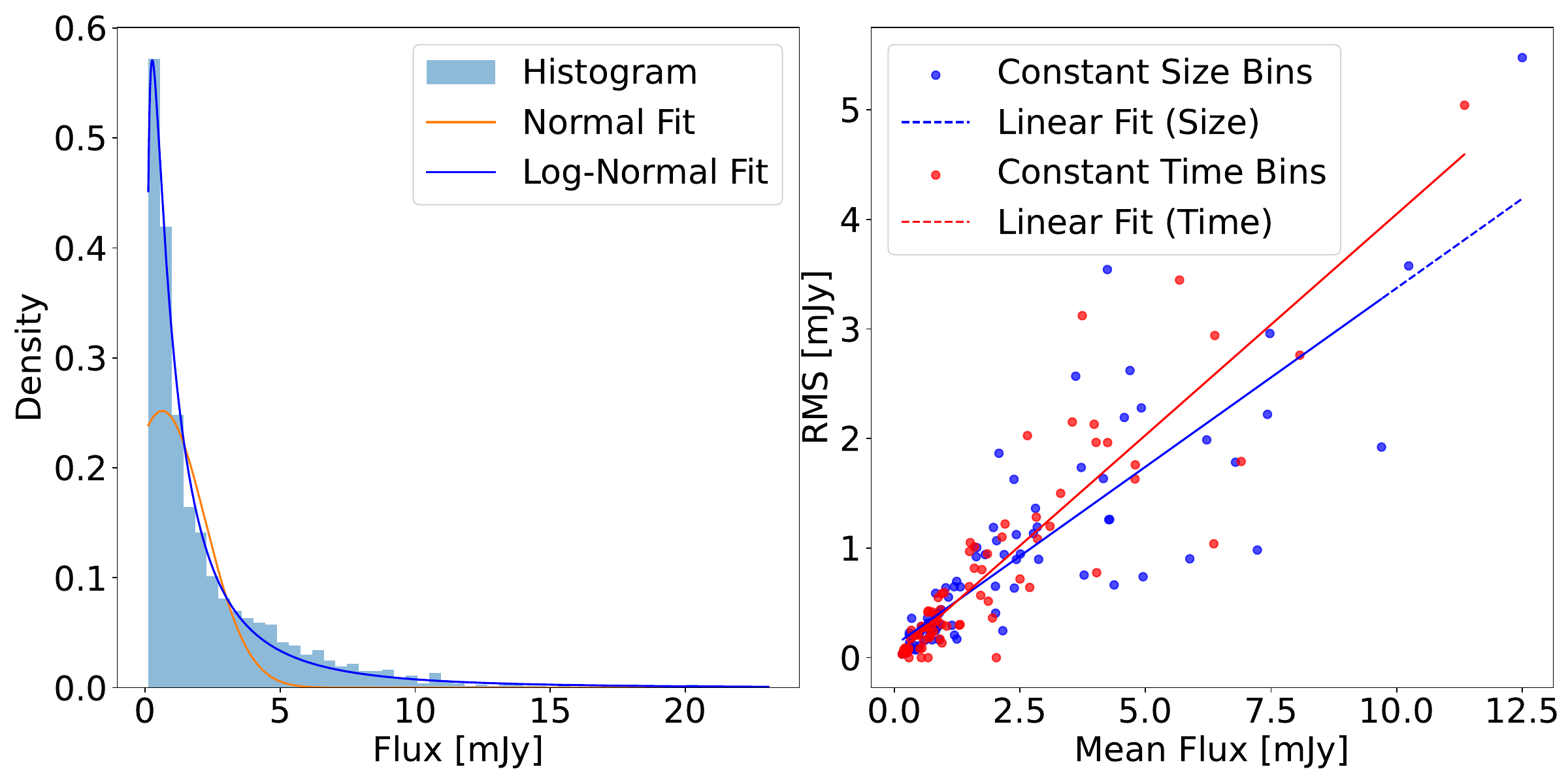}
\caption{Left panel: Flux distribution from $\sim$12 years of R-band observations. The histogram (blue bars) shows the flux counts. The red curve is the best-fitting normal function, and the blue curve is the best-fitting log-normal function. Right panel: RMS (i.e., the square root of the excess variance) as a function of the mean flux, computed in (1) bins containing an equal number of data points (blue markers), and (2) bins of fixed time length (red markers). The solid blue and red lines indicate the best linear fits for each binning scheme.}
\label{fig: fluxPDF_fluxRMS_relations}
\end{figure}

\paragraph{Flux distribution.} The distribution of the blazar radiation flux can provide clues about the underlying nature of the radiation process. A normal flux distribution typically suggests additive processes, where the total radiation arises from the sum of contributions from multiple independent sources (e.g., independent emission regions or individual shocks within the blazar). In contrast, a log-normal flux distribution implies multiplicative processes, where the radiation is shaped by a chain of factors that multiplicatively influence the total flux (e.g., variability driven by cascading turbulent processes, relativistic beaming, or interactions in a jet). In some blazars, neither of these distributions is suitable to describe the flux distribution \citep[see, e.g.,][where the flux distribution in the $\gamma$ domain suggests the existence of two log-normal distributions]{2018RAA....18..141S}.
\\
\indent We performed a flux distribution analysis for the entire light curve across all filters. As an example, the left panel of Figure \ref{fig: fluxPDF_fluxRMS_relations} displays the flux distribution in the R band\footnote{The histogram binning was performed using the \texttt{auto} method in NumPy, which selects the bin width based on the minimum of the Sturges and Freedman–Diaconis rules. This approach provides a data-driven, statistically motivated binning scheme.}. Prior to creating the histogram, the data were averaged into one-day bins to reduce noise. The histogram was then fit with both normal and log-normal functions, represented by the smooth red and blue lines, respectively. The reduced $\chi^2$ values for the normal and log-normal fits indicate that the flux distribution is better described by the log-normal function in all filters. Specifically, the reduced $\chi^2$ values for the normal (log-normal) fits to the distributions for the B, V, R, and I bands are 24.43 (4.54), 20.62 (2.85), 29.76 (3.10), and 11.01 (2.13), respectively.
\\
\\
\indent The results of the histogram-based analysis depend on the choice of bin size. To address this, we performed fitting on the unbinned flux values using the maximum likelihood estimation method. To evaluate the goodness-of-fit, we utilized the Bayesian information criterion (BIC) and Akaike information criterion (AIC) statistical measures. The lower values of both BIC and AIC for the log-normal function compared to the normal function further confirm that the flux distribution is better described by a log-normal model.
\\
\indent In addition to the previous analysis, we also performed the Kolmogorov-Smirnov (KS) test to further assess the goodness-of-fit for the flux distributions. The KS test compares the empirical cumulative distribution of the data with theoretical distributions (normal and log-normal), providing two key statistics: \(D\), the maximum deviation between the observed and expected cumulative distributions, and \(p\), the p-value, which indicates the significance of the fit. The results of the KS test are given in Table \ref{tab:KS_test}. These results reinforce the conclusion that the flux distribution of the blazar is better described by a log-normal distribution in all observed filters.
\begin{table}[ht]
\centering
\caption{Kolmogorov-Smirnov test statistics and p values for normal and log-normal distributions in different filters}
\label{tab:KS_test}
\begin{tabular}{lcc|cc}
\hline
Filter & \multicolumn{2}{c|}{Normal distribution} & \multicolumn{2}{c}{Log-normal distribution} \\
       & D & p value & D & p value \\
\hline
B  & 0.36 & $2.45 \times 10^{-75}$  & 0.05 & $7.23 \times 10^{-2}$  \\
V  & 0.36 & $6.23 \times 10^{-96}$  & 0.04 & $1.98 \times 10^{-1}$  \\
R  & 0.37 & $1.46 \times 10^{-209}$ & 0.04 & $1.48 \times 10^{-2}$  \\
I  & 0.37 & $1.18 \times 10^{-89}$  & 0.04 & $1.96 \times 10^{-1}$  \\
\hline
\end{tabular}
\end{table}
\\
\indent Finally, we give the parameters that best fit a log-normal distribution of the form
\[ f(x) = \frac{1}{x s \sqrt{2 \pi}} \exp{\left[ -\frac{(ln(x) - m)^2}{2 s^2} \right]}, \]
where m and s are the mean location (in the natural logarithm space) and shape parameter (standard deviation in the natural logarithm space):
\begin{flushleft}
B: $m = -0.05 \pm 0.10$, $s = 1.24 \pm 0.10$ \\
V: $m = 0.18 \pm 0.06$, $s = 1.17 \pm 0.06$ \\
R: $m = 0.20 \pm 0.04$, $s = 1.24 \pm 0.04$ \\
I: $m = 0.81 \pm 0.07$, $s = 1.17 \pm 0.08$ \\
.\end{flushleft}

\begin{figure}
\centering    
\includegraphics[width=0.5\textwidth,keepaspectratio=true]{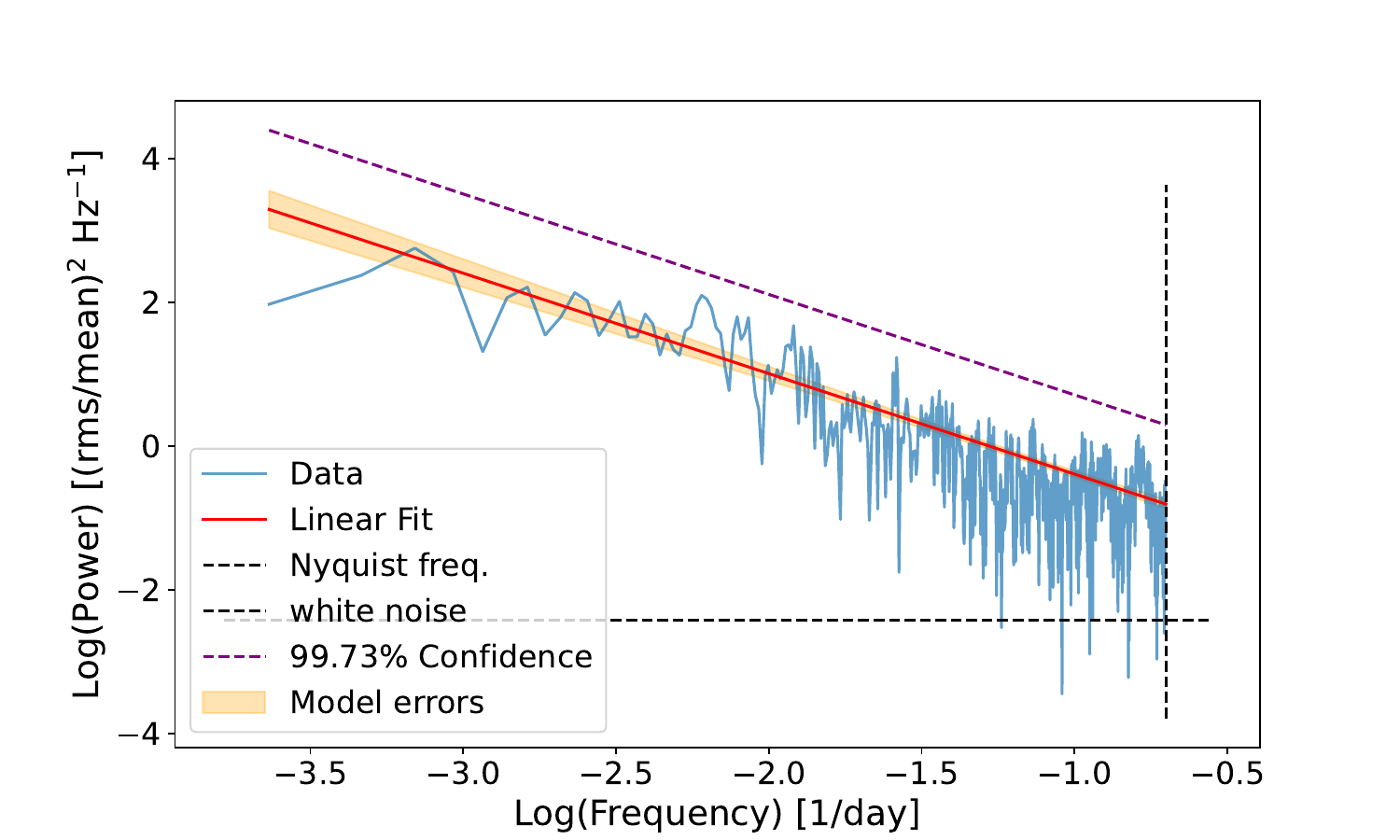}
\caption{R-band periodogram of Ton 599.}
\label{fig: PSD_Vaughen(2005)_Rband}
\end{figure}

\paragraph{RMS-flux relation.} To further characterize the flux distribution and variation, we analyzed the dependence of the degree of flux variation on the mean radiation flux (known as the RMS-flux relation). We binned the data in two ways to calculate both the degree of variation and the mean flux within each bin: a) using fixed time intervals of 50 days, and b) ensuring that each bin contained exactly 20 data points. To quantify the amplitude of the flux variation in each bin, we employed the excess variance $\sigma_{XS} = \sqrt{S^2 - \left<\sigma^2_{err}\right>}$, where $S^2$ is the variance of flux in the light curve and $<\sigma^2_{err}>$ the average of squared measurement error.
\\
\indent The right panel of Figure \ref{fig: fluxPDF_fluxRMS_relations} shows the RMS-flux relation for the two binning modes for the R-band light curve. Red dots correspond to binning with a fixed number of data, and blue dots correspond to binning with a fixed time interval. We fit a straight line using the least-squares method; the fit is shown in the figure as a straight line with the same color as the corresponding data. The analysis shows a significant positive correlation for all filters, and the results are presented in Table \ref{tab: RMS-flux relation}.
\\
\begin{table*}[ht]
\centering
\caption{Linear regression parameters of the RMS–flux relation in different filters}
\label{tab: RMS-flux relation}
\begin{tabular}{lccc|ccc}
\hline
Filter & \multicolumn{3}{c|}{Size correlation} & \multicolumn{3}{c}{Time correlation} \\
       & Slope $\pm$ Error & Intercept $\pm$ Error & $r$ (p value) & Slope $\pm$ Error & Intercept $\pm$ Error & $r$ (p value) \\
\hline
B  & $0.39 \pm 0.05$ & $0.09 \pm 0.11$ & $0.81$ ($2.4 \times 10^{-9}$)  & $0.36 \pm 0.03$ & $-0.02 \pm 0.06$ & $0.80$ ($6.2 \times 10^{-16}$) \\
V  & $0.29 \pm 0.03$ & $0.21 \pm 0.09$ & $0.79$ ($6.1 \times 10^{-12}$) & $0.30 \pm 0.03$ & $0.06 \pm 0.08$  & $0.74$ ($1.2 \times 10^{-13}$) \\
R  & $0.25 \pm 0.02$ & $0.17 \pm 0.07$ & $0.78$ ($3.0 \times 10^{-25}$) & $0.41 \pm 0.02$ & $0.00 \pm 0.06$  & $0.90$ ($2.6 \times 10^{-29}$) \\
I  & $0.26 \pm 0.04$ & $0.54 \pm 0.21$ & $0.69$ ($3.3 \times 10^{-7}$)  & $0.29 \pm 0.03$ & $0.20 \pm 0.14$  & $0.75$ ($2.5 \times 10^{-13}$) \\
\hline
\end{tabular}
\end{table*}
\\
\paragraph{PSD distribution.} The power spectral density (PSD) of blazars in the optical domain often follows a power law, $P(\nu)\propto \nu^{-\beta}$, with typical slopes of $\beta\sim 1-3$. This indicates that flux variability of blazars is a correlated colored-noise stochastic process, with the amplitude of flux variations decreasing toward shorter timescales (specifically, $\beta \sim 0$ is a uncorrelated white noise process, $\beta \sim 1$ is a correlated pink (or flickering) noise process, $\beta \sim 2$ is a red (or random-walk) noise process, $\beta \geq 3 $ is a black noise process). Determining the slope and normalization of the PSD, as well as the inflection points, is of great importance because they carry important information about the system in which variations occur, such as the size of the emission region or the timescale of particle cooling in the emission region.
\\
\indent We adopted the method described by \citet{2005A&A...431..391V} to compute the periodogram of our light curves. Concretely, if $x_j$ ($j=1,\dots,N$) denotes the time series of flux values, then we define the periodogram, $P(\nu_k)$, as
\[
P(\nu_k) \;=\; \frac{2\,T}{\langle x\rangle^2\,N^2}\,\bigl|\mathrm{DFT}(\nu_k)\bigr|^2,
\]
where $\mathrm{DFT}(\nu_k)$ is the discrete Fourier transform of the light curve, $T \equiv \frac{N (t_N - t_1)}{N-1}$ is the effective duration of the time series, and $\langle x\rangle$ is the mean flux. We computed the DFT at frequencies $\nu_k = k / T$ ($k=1,\dots,N/2$), so that $\nu_{\min}=1/T$ and $\nu_{\mathrm{Ny}}=2/(N\,T)$ is the Nyquist frequency. By integrating the positive part of this periodogram, one recovers the fractional variability (rms$/\langle x\rangle$).
\\ 
\indent Because our data are not uniformly sampled, we interpolated the data onto a uniform grid with a time step of $T/N$. Next, we subtracted the mean flux (so that the time series has zero mean) and then applied a Hanning window to reduce leakage effects (see \cite{2014MNRAS.445..437M} for the importance of centering the light curve and using a Hanning window before calculating the periodogram). The resulting periodogram, $\log P(\nu_k)$ versus $\log \nu_k$, for the R bandpass is shown in Fig.~\ref{fig: PSD_Vaughen(2005)_Rband}. The vertical dashed line corresponds to the Nyquist frequency. The horizontal dashed line corresponds to the power that comes from the uncertainty of measurement, calculated as $P_{noise} = (2 T)/(N \left<{x}\right>^2) (\sum_{j=1}^{N} \sigma^2 / N)$, where $\sigma$ is the flux error.
\\
\indent Following the method of \citet{2005A&A...431..391V}, we fit the raw (unbinned) periodogram with a linear function in log--log space using the least-squares method, obtaining both the best-fit parameters and their uncertainties, as well as model-based error estimates. In Figure~\ref{fig: PSD_Vaughen(2005)_Rband}, the red line represents the best-fitting power-law model, while the shaded region indicates the range of model errors around that fit. For the R band, we find a power law slope of $-1.40 \pm 0.05$. We also performed a KS test on the ratio of the observed periodogram to the model, confirming that it follows a $\chi^2$ distribution with two degrees of freedom ($\chi^2_2$). The KS statistic (and corresponding $p$ value) is $2.37 \times 10^{-2} (7.12 \times 10^{-1})$.
\\
\indent We calculated the 99.73\% confidence limit by evaluating the quantity
\[
\gamma_{\epsilon} \;=\; -2 \,\ln\Bigl[\,1 \;-\;\bigl(1 - \epsilon_{n'}\bigr)^{1/n'}\Bigr],
\]
where $\epsilon$ represents the false-alarm probability, and $n' = n - 1$ indicates that the Nyquist frequency is excluded from the calculation. We then added $\log(\gamma_{\epsilon}/2)$ to the periodogram to define the threshold level. The 99.73\% confidence limit is drawn as dashed line above the periodogram, and it shows that there is no statistically significant peak indicating periodicity in our light curve. We obtained a similar result for the light curves for the B, V, and I bands. Here we list only the slopes of the obtained periodograms: $-1.79 \pm 0.07$, $-1.53 \pm 0.06$ and $-1.69 \pm 0.07$, respectively.

\subsection{Intranight and intraday flux variations}

\begin{figure*}
\centering    \includegraphics[width=\textwidth,keepaspectratio=false]{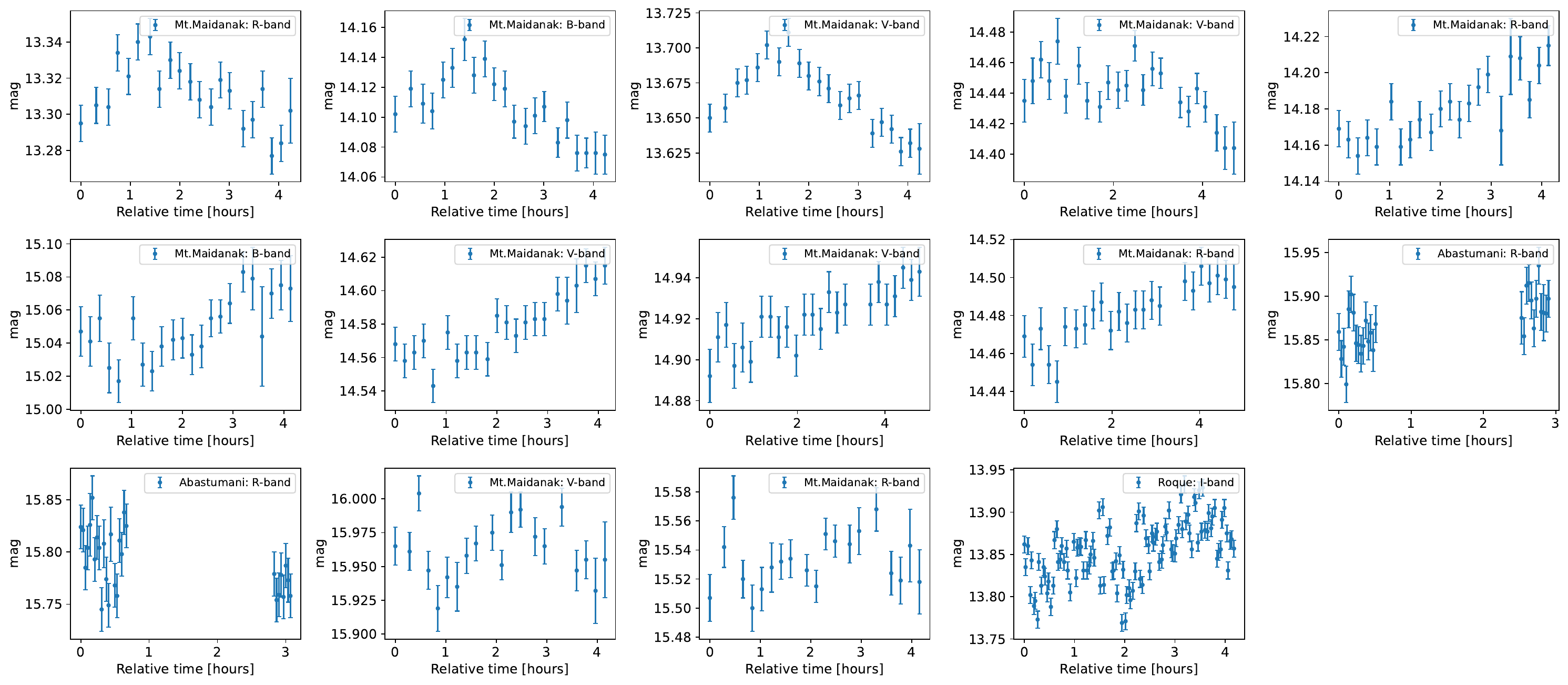}	
\caption{Intranight light curves displaying statistically significant variability based on the $\chi^2$ test. The vertical axis shows the measured magnitude, the legend indicates the filter used for each observation, and the horizontal axis represents the relative time in hours.}
\label{fig: IDV_variability}
\end{figure*}

\indent In order to study the shortest variations in brightness of the blazar Ton 599, we extracted the datasets that were taken from the same observatory within a single observing night. For more precise statistical analysis, we considered only datasets with more than 2 hours of observation and more than 20 data points. With these criteria, 31 light curves were extracted.
\\
\indent We used a \(\chi^2\) test to determine whether the light curves exhibit statistically significant variability on a given night. The \(\chi^2\) statistic was computed as
\begin{equation}
\chi^2 \;=\; \sum_{i=1}^{N} \frac{(O_i - E_i)^2}{\sigma_i^2},
\end{equation}
where \(O_i\) is the observed flux, \(E_i\) is the expected value (here, the mean flux), \(\sigma_i\) is the measurement uncertainty for the \(i\)th observation, and \(N\) is the total number of data points. We then compared the resulting \(\chi^2\) value to a \(\chi^2\) distribution with \(N - 1\) degrees of freedom to obtain a p-value (\(p\)), which reflects the likelihood that the observed variations are solely due to measurement errors. We adopted \(p < 0.05\) as the threshold to reject the null hypothesis of a constant source flux. Consequently, \(p < 0.05\) indicates significant variability, whereas \(p \ge 0.05\) suggests that any apparent flux changes could be attributed to random noise rather than genuine variability. As a result of the test, we found statistically significant variability in 14 intranight light curves across multiple filters, corresponding to 8 distinct nights of observation..
\\
\indent We adopted the fractional variability measure from \citet{Peterson2001} to characterize the amplitude of flux variations in the intranight light curves. It is defined by
\begin{equation}
F_{\mathrm{var}} \;=\; \frac{\sqrt{\sigma^{2} - \delta^{2}}}{\langle F \rangle},
\label{eq:F_var}
\end{equation}
where \(\sigma^{2}\) is the sample variance of the fluxes, \(\delta^{2}\) is the mean square measurement error, and \(\langle F \rangle\) is the mean flux. Errors for $F_{\mathrm{var}}$ were calculated as follows \citep[e.g.,][]{2020ApJ...891..120B}:
\begin{equation}
\sigma_{F_{\text{var}}} = \sqrt{ F_{\text{var}}^2 + \sqrt{\frac{2}{N} \frac{\langle \sigma_{\text{err}}^2 \rangle^2}{\langle F \rangle^4} + \frac{4}{N} \frac{\langle \sigma_{\text{err}}^2 \rangle}{\langle F \rangle^2} F_{\text{var}}^2}} - F_{\text{var}}.
\end{equation}
This definition accounts for measurement errors, thereby enabling the straightforward identification of light curves in which measurement uncertainties dominate the sample variance.
\\
\indent Table \ref{tab: IDV variability} shows the result of the $\chi^2$ test (column 4) and the $F_{var}$ measurement (column 5) for observation nights when the blazar was variable according to the $\chi^2$ test. Considering the observation dates (column 2) and filters (column 3), the source was variable on 8 nights, with a total of 14 intranight light curves showing significant variability..
\\
\begin{table*}[t]
\tiny
\caption{Intranight variability of Ton~599 from optical monitoring}
\label{tab: IDV variability}
\centering
\begin{minipage}{\textwidth}  
\centering
\resizebox{\columnwidth}{!}{ 
\begin{tabular}{lllrllrr}
\hline
 Observatory   & Date       & Filter   &   p value & $F_{var} \pm \sigma \; [\%]$   & $\tau_{var} \pm \sigma \; [h]$          &   R [cm] &   B [Gauss] \\
\hline
 Mt.Maidanak (60cm)  & 2017-12-10 & R        &   6.5e-06 & 1.3 $\pm$ 0.2                  & 5.5 $\pm$ 2.1                     &  1.9e+17 &        0.19 \\
 Mt.Maidanak (60cm)   & 2017-12-10 & B        &   6.8e-07 & 1.7 $\pm$ 0.2                  & 12.5 $\pm$ 7.4                   &  4.3e+17 &        0.14 \\
 Mt.Maidanak (60cm)   & 2017-12-10 & V        &   6.8e-14 & 2.0 $\pm$ 0.2                  & 9.7 $\pm$ 6.5                    &  3.4e+17 &        0.15 \\
 Mt.Maidanak (60cm)   & 2017-12-12 & V        &   2.8e-03 & 1.3 $\pm$ 0.2                  & 5.6 $\pm$ 2.9                    &  2.0e+17 &        0.19 \\
 Mt.Maidanak (60cm)   & 2017-12-15 & R        &   3.6e-05 & 1.3 $\pm$ 0.2                  & 4.9 $\pm$ 3.4                    &  1.7e+17 &        0.19 \\
 Mt.Maidanak (60cm)   & 2017-12-15 & B        &   7.3e-03 & 1.1 $\pm$ 0.3                  & 6.8 $\pm$ 4.6                    &  2.3e+17 &        0.17 \\
 Mt.Maidanak (60cm)   & 2017-12-15 & V        &   4.4e-08 & 1.5 $\pm$ 0.2                  & 7.5 $\pm$ 3.9                    &  2.6e+17 &        0.17 \\
 Mt.Maidanak (60cm)   & 2017-12-19 & V        &   1.1e-02 & 0.9 $\pm$ 0.2                  & 12.2 $\pm$ 7.8                   &  4.2e+17 &        0.14 \\
 Mt.Maidanak (60cm)   & 2017-12-19 & R        &   7.0e-04 & 1.1 $\pm$ 0.2                  & 7.0 $\pm$ 3.6                    &  2.4e+17 &        0.17 \\
 Abastumani    & 2018-01-02 & R        &   9.6e-04 & 1.9 $\pm$ 0.4                  & 0.4 $\pm$ 0.1                           &  1.5e+16 &        0.44 \\
 Abastumani    & 2018-01-03 & R        &   2.7e-03 & 1.9 $\pm$ 0.4                  & 0.5 $\pm$ 0.3                           &  1.9e+16 &        0.41 \\
 Mt.Maidanak (60cm)   & 2018-01-14 & V        &   1.4e-03 & 1.4 $\pm$ 0.4                  & 3.6 $\pm$ 1.2                    &  1.2e+17 &        0.22 \\
 Mt.Maidanak (60cm)   & 2018-01-14 & R        &   1.2e-02 & 1.2 $\pm$ 0.4                  & 3.6 $\pm$ 1.3                    &  1.3e+17 &        0.22 \\
 Roque         & 2018-05-12 & I        &   0.0e-00 & 3.2 $\pm$ 0.1                  & 0.3 $\pm$ 0.0                           &  9.8e+15 &        0.50 \\
\hline
\end{tabular}
}
\end{minipage}
\tablefoot{Columns list the observatory, date, and filter of observation. 
$p$ is the probability from the $\chi^{2}$ variability test. 
$F_{\mathrm{var}}$ is the fractional variability amplitude with error. 
$\tau_{\mathrm{var}}$ is the minimum variability timescale in hours. 
$R$ is the radius of the emitting region in centimeters, and $B$ is the magnetic field strength in Gauss.}

\end{table*}
\\
\indent Figure \ref{fig: IDV_variability} shows the intranight light curves for which the $\chi^{2}$ test showed to be variable. In the reading direction, the subplots are arranged in the same order as in the Table \ref{tab: IDV variability}. The timescale is expressed in hours relative to the starting time of observation for a given night. In the legends of each subplot, we give the filter and the name of the observatory.
\\
\\
\indent For each light curve listed in Table \ref{tab: IDV variability}, we calculated the variability timescale following the prescription of \citet{1974ApJ...193...43B}:
\[
\tau_{ij} = \frac{\Delta t}{\ln \bigl(F_i / F_j\bigr)},
\]
where \(\Delta t\) is the time difference between two flux measurements \(F_i\) and \(F_j\). We calculated \(\tau_{ij}\) for each pair \((i,j)\) of measurements that satisfies
\(\lvert F_i - F_j\rvert > \bigl(\sigma_{F_i} + \sigma_{F_j}\bigr)\). The characteristic timescale of variability is then taken to be the minimum of all computed \(\tau_{ij}\) values. We derived the corresponding uncertainty via standard error propagation from the flux measurement errors. The timescales of variability for each epoch are given in column 6 of Table \ref{tab: IDV variability}.
\\
\indent The upper limit on the size of the emission region can be estimated using the relation
\[
R \;\le\; c\,\tau_{\mathrm{min}}\,\frac{\delta}{1 + z},
\]
where \( \delta \) is the Doppler boosting factor. For our calculations, we adopted a mean Doppler factor of \( \delta = 23.1 \), derived from multiple estimates reported in the literature. Specifically, \cite{2017ApJ...846...98J} obtained \( \delta = 11.8 \) based on Very Long Baseline Array (VLBA) observations, utilizing measurements of the apparent speeds of moving knots and variability timescales from light curves. \cite{2010A&A...512A..24S} determined \( \delta = 28.2 \) using a combination of VLBA and Metsähovi data. \cite{2018ApJS..235...39C} estimated \( \delta = 29.3 \) from broadband spectral energy distribution modeling. The adopted value represents an average of these independent estimates, providing a robust constraint on the size of the emission region.
\\
\indent To estimate the magnetic field strength, we applied the condition that the observed minimum variability timescale, derived from intranight observations, must be at least as long as the synchrotron cooling timescale \citep{2008ApJ...672...40H}:
\[
t_{\mathrm{syn}} \;\approx\; 4.75 \times 10^{2}\,\biggl(\frac{1 + z}{\delta\,\nu\,B^{3}}\biggr)^{\!\!0.5},
\]
where \(\nu\) is the observation frequency in Hz, \(t_{\mathrm{syn}}\) is in days, and \(B\) is in Gauss. The results of the analysis for the size of the emission region and B are given in columns 7 and 8 of Table \ref{tab: IDV variability}.
\\
\indent \cite{2024A&A...686A.228O} demonstrated that the flux distributions derived from TESS observations on IDV timescales are well fit by a Gaussian model. They argue that the success of the Gaussian representation suggests that the rapid variability is generated in a compact region of the jet through linear (additive) processes rather than by disk-related mechanisms. In our study, we applied two normality tests - the Shapiro–Wilk \citep{Shapiro} and Anderson–Darling \citep{Anderson} tests - to the light curves presented in Table~\ref{tab: IDV variability}. Our results indicate that the IDV flux distributions in all cases are consistent with a Gaussian profile.

\subsection{Long-term variability}

\begin{figure}
	\centering    \includegraphics[width=0.5\textwidth,keepaspectratio=false]{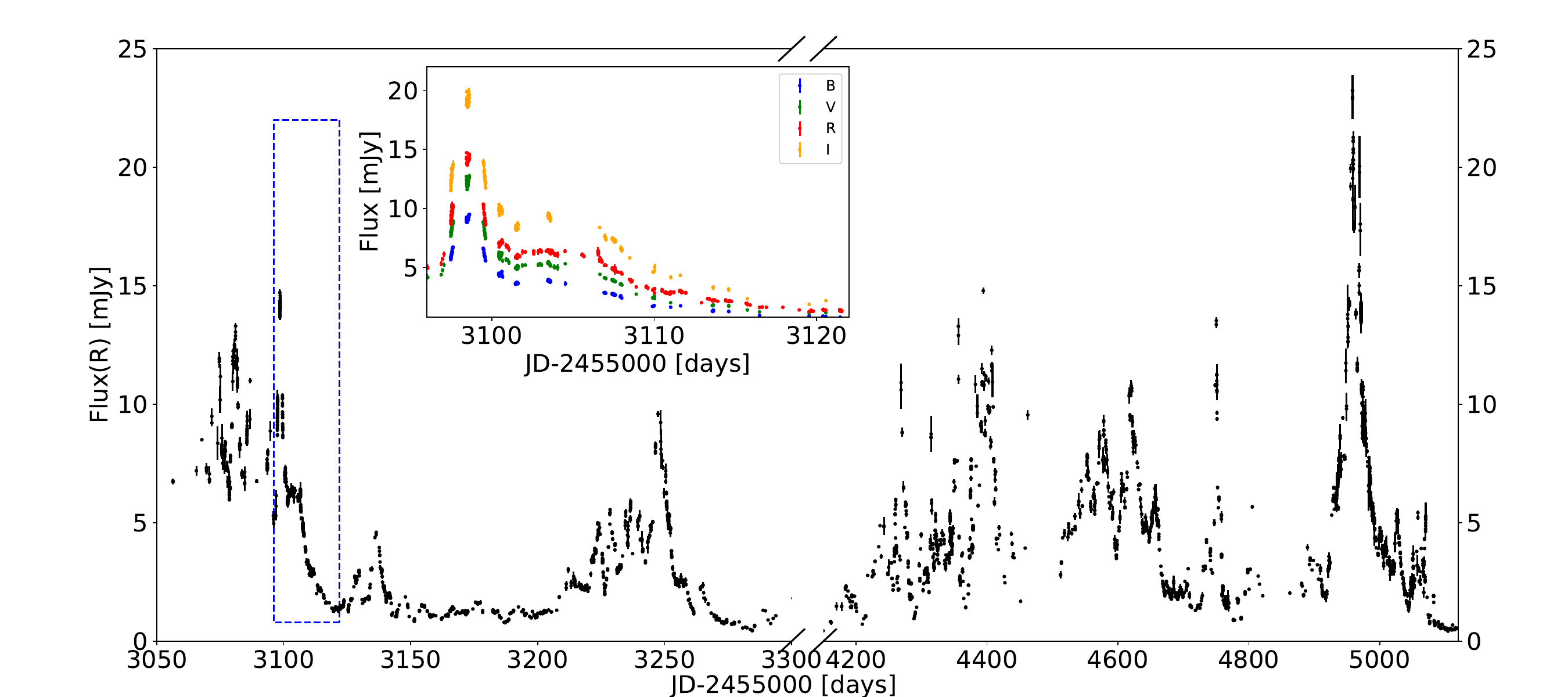}	
	\caption{Zoomed-in R-band light curve from JD-2455000=3050 to 3300 (Period II) and from JD-2455000=4150 to 5120 (Period III) in the R band. Complex structure of the outbursts is visible. The inset highlights the second-largest peak, well sampled in all filters.}
	\label{fig: STV_outburst}
\end{figure}

\indent As can be seen from Figure \ref{fig: F-PD-EVPA_vs_JD}, over a period of 12 years, the Ton 599 blazar goes through alternating phases of weak and strong activity. The average flux density for the entire time series in the BVRI filters is 2.50, 3.16, 3.31, and 5.73 mJy, respectively. For the same order of filters, the flux density range (minimum–maximum) is 0.11-9.49, 0.13-12.75, 0.11-23.25, and 0.14-20.39 mJy, respectively. We note here that, due to the difference in sampling in the BVRI filters, the maximum flux peaks do not coincide in time. For instance, in the R filter, the maximum flux is recorded around JD-2455000 = 4958.5, which is not at all covered by the observations in the other filters.
\\
\indent The time interval from November 2011 to September 2017, i.e., from JD-2455000=900 to 3000, is marked by several average-intensity flares (Period I). The maximum fluxes in this period are 3.82, 5.42, 6.49, and 10.10 mJy for the BVRI filter, which, due to different sampling, do not occur in the same epoch.
\\ 
\indent This period is followed by two large outbursts that occur in about 400 days from about JD-2455000=3000 to 3400 (Period II). Their structure is much more complex than it can be seen in Figure \ref{fig: F-PD-EVPA_vs_JD} and consists of a series of smaller-larger flares (see Figure \ref{fig: STV_outburst}). The peak of the first outburst, which in all filters occurs nearly simultaneously around JD-2455000=3098 (see Table \ref{tab: LDV_summary}), reaches the value of 9.48, 12.75, 14.71, and 19.89 mJy for the BVRI filters, respectively (see the inset plot of the Figure \ref{fig: STV_outburst}). According to the R-band light curve, this peak is the second largest for the entire observation period. Assuming that the spectrum can be described with a power law of the form $F_{\nu} \sim \nu^{-\alpha}$, where $\alpha$ is the spectral index, we calculated $\alpha$ for the 3098 peak using the relation 
\begin{equation}
\alpha = {{log(F_{1}/F_{2})}\over{log(\nu_{1}/\nu_{2})}},
\label{eq: alpha_from_flux}
\end{equation}
where the subscripts 1 and 2 denote any combination of filters. By averaging spectral indices obtained from all possible combinations of filters (BV, BR, BI, VR, VI, RI), the mean value is $\alpha = 1.2 \pm 0.2$, where the error is given as the standard deviation of individual $\alpha$ values. After correction to the cosmological redshift, using equation $F_{\text{corr}} = F(1 + z)^{3+\alpha}$ \citep{2018A&A...620A.185N}, the logarithms of luminosity for the BVRI filters are 48.21, 48.24, 48.23, and 48.27 erg/s, respectively, which is on average for the optical equal to $log(\nu L_{\nu}) = 48.24 \pm 0.02$ [erg/s]. It is worth noting that the maximum flux density for the entire time series in the R band is 23.25 mJy, which is about 1.6 times greater than this peak value.
\\
\noindent The peak of the second outburst in period II, which also occurs almost simultaneously (within one day around  JD-2455000=3248), is 5.09, 7.07, 9.59, and 14.12 mJy for BVRI, respectively. With the same procedure as for the first peak, for this peak we find that the logarithm of the luminosity is 48.06, 48.11, 48.17, and 48.25 erg/s in the BVRI filters; that is, in the average $log(\nu L_{\nu}) = 48.15 \pm 0.07$ [erg/s].
\\
\\
\indent After these two outbursts, there follows a period of about 2.2 years of relatively weak activity, marked by only two smaller flares (maximum flux in the R filter about 2.2 mJy). This phase is interrupted by an active phase lasting about 2.8 years (from about JD-2455000=4200 to the end of the time series; Period III) which is marked by a series of large flares. This period also saw the biggest peak in the blazar light curve, with flux density 23.25 mJy in the R band, which is a record for the period we are analyzing. As we mentioned earlier, the maximum flux of 23.25 mJy in the R band was not observed in other filters, so it is not possible to determine the luminosity as for the peaks in the period II. However, we can make an estimate using the average spectral indices that we calculate in Sect. \ref{sec: Color analysis}. The mean spectral indices for the B-V, B-R, B-I, V-R, V-I, and R-I color indices (see Figure \ref{fig: color_VS_mag}) are 1.38, 1.30, 1.43, 1.21, 1.45, and 1.64, respectively, from which it follows that the mean luminosity for the highest peak in the R band is $log(\nu L_{\nu}) = 48.48 \pm 0.03$ [erg/s], where the error is calculated as the standard deviation of the luminosity values calculated with different mean $\alpha$.
\\ 
\indent Table \ref{tab: LDV_summary} summarizes some of the variability parameters mentioned above. The table is divided into four large groups, which present the parameters for the entire time series, period I, period II, and period III, respectively, as indicated in their headers. Each of these groups list four parameters: the JD-2455000 of the maximum flux density, the average flux density in mJy, the maximum flux density in mJy, and the amplitude of variation calculated using equation \ref{eq:F_var} in percentage.
\\ 
\indent Figure \ref{fig: plot_outbursts_4x3} provides an overview of the entire dataset,  divided into 12 time segments, each highlighted in light blue in the top panel. The lower panels offer a closer view of each segment, arranged sequentially from left to right.
\\
\begin{table*}
\centering
\caption{Long-term variability of Ton~599 in different time segments}
\label{tab: LDV_summary}
\resizebox{\textwidth}{!}{
\begin{tabular}{c|cccc|cccc|cccc|cccc}
\hline
 & \multicolumn{4}{c|}{Whole (870-5200)} & \multicolumn{4}{c|}{Period I (900-3000)} & \multicolumn{4}{c|}{Period II (3000-3400)} & \multicolumn{4}{c}{Period III (4200-5200)} \\
 & $Epoch_{max}$ & $\left<F\right>$ & $F_{max}$ & $F_{var}$ & $Epoch_{max}$ & $\left<F\right>$ & $F_{max}$ & $F_{var}$ & $Epoch_{max}$ & $\left<F\right>$ & $F_{max}$ & $F_{var}$ & $Epoch_{max}$ & $\left<F\right>$ & $F_{max}$ & $F_{var}$ \\
\hline
B & 4750.35 & 2.50 & 9.49 & 84 & 1101.30 & 0.80 & 3.82 & 74 & 3098.62 & 3.25 & 9.48 & 74 & 4750.35 & 3.15 & 9.49 & 49 \\
V & 3098.61 & 3.16 & 12.75 & 85 & 2746.59 & 1.07 & 5.42 & 83 & 3098.61 & 4.43 & 12.75 & 72 & 4394.33 & 4.10 & 12.42 & 46 \\
R & 4958.51 & 3.31 & 23.25 & 99 & 2746.57 & 1.24 & 6.49 & 80 & 3098.44 & 4.24 & 14.71 & 75 & 4958.51 & 5.24 & 23.25 & 67 \\
I & 4394.32 & 5.73 & 20.39 & 77 & 2746.58 & 1.79 & 10.10 & 79 & 3098.58 & 7.59 & 19.89 & 62 & 4394.32 & 7.46 & 20.39 & 43 \\
\hline
\end{tabular}
}
\tablefoot{Columns correspond to the whole observing period (JD–2455000 $\sim$ 870–5200) 
and to three sub-periods (I: JD–2455000 $\sim$ 900–3000, II: JD–2455000 $\sim$ 3000–3400, III: JD–2455000 $\sim$ 4200–5200). 
For each segment we list the epoch of maximum flux density ($Epoch_{max}$, in JD–2455000), 
the average flux density ($\langle F \rangle$, in mJy), the maximum flux density 
($F_{max}$, in mJy), and the fractional variability amplitude ($F_{var}$, in percent), 
calculated using Equation \ref{eq:F_var}.}
\end{table*}

\subsection{Color analysis}\label{sec: Color analysis}

\begin{figure}
	\centering
	\resizebox{9cm}{!}{\includegraphics{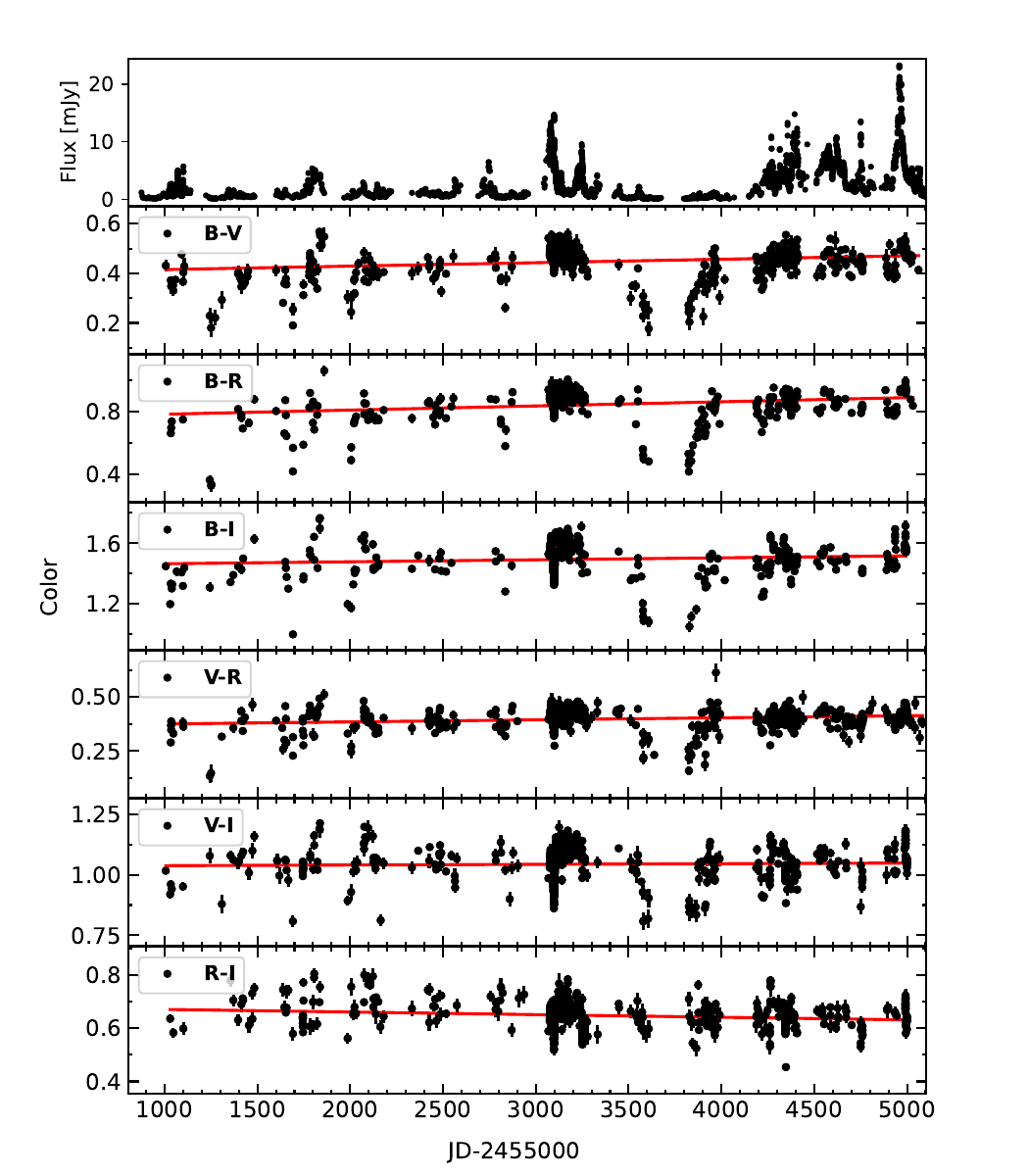}}
	\vspace{-3mm}
	\caption{Evolution of the color indices with time. The first panel from the top shows the light curve in the $R$ filter. The rest of the panels present the CI evolution. The red line is the best fit of a linear function to the data. Parameters of the linear function are provided in the Table \ref{tab: color_behaviour}.}
	\label{fig: color_VS_time_wLightcurve}
	\vspace{-1mm}
\end{figure}

\begin{figure}
	\centering
	\includegraphics[height=11cm,width=\columnwidth]{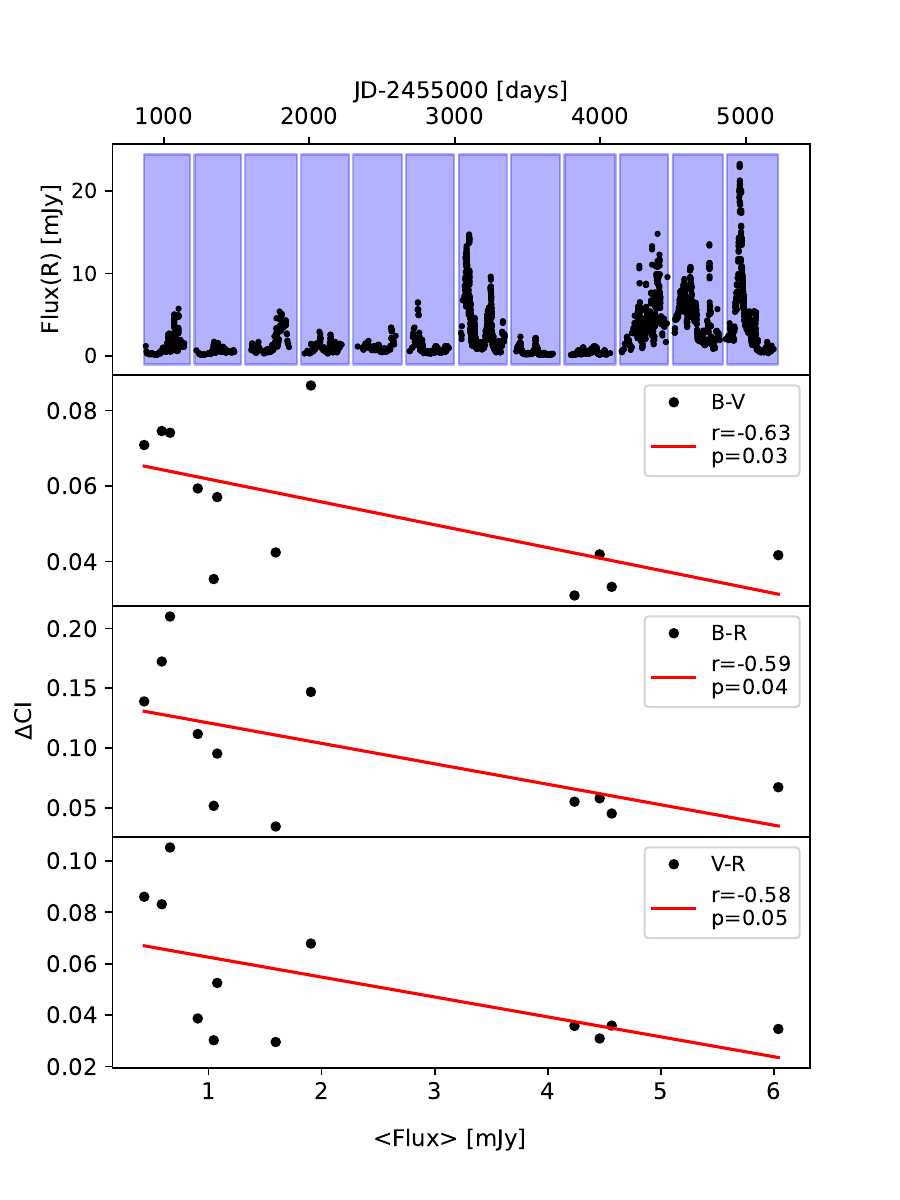}
	\vspace{-3mm}
	\caption{Correlation between variation in the color indices and the mean flux density in 12 segments (shaded in the top R-band light curve). Red lines show the best linear fits.}
	\label{fig: dCI_VS_aveFlux}
	\vspace{-1mm}
\end{figure}

\begin{figure}
	\centering
	\resizebox{9cm}{!}{\includegraphics{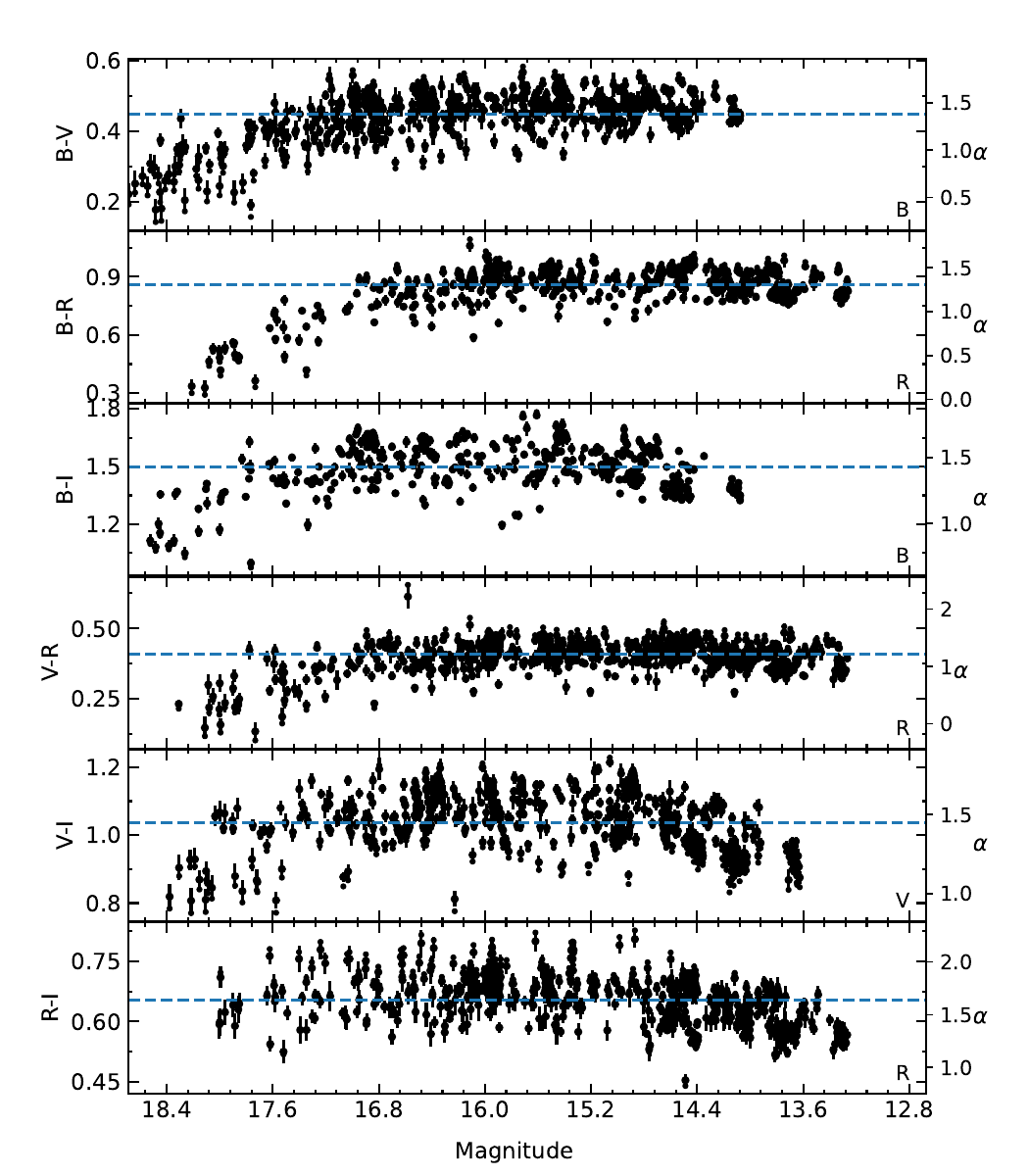}}
	\vspace{-3mm}
	\caption{Dependence of the color indices and spectral slopes on brightness. The dashed blue line corresponds to the mean value of the color indices. The right y axis shows the spectral slope, $\alpha$, which was calculated under the assumption that the optical flux can be described by the relation $F_{\nu} \sim \nu^{-\alpha}$ (see the text).}
	\label{fig: color_VS_mag}
	\vspace{-1mm}
\end{figure}

\begin{figure}
	\centering
	\resizebox{9cm}{!}{\includegraphics{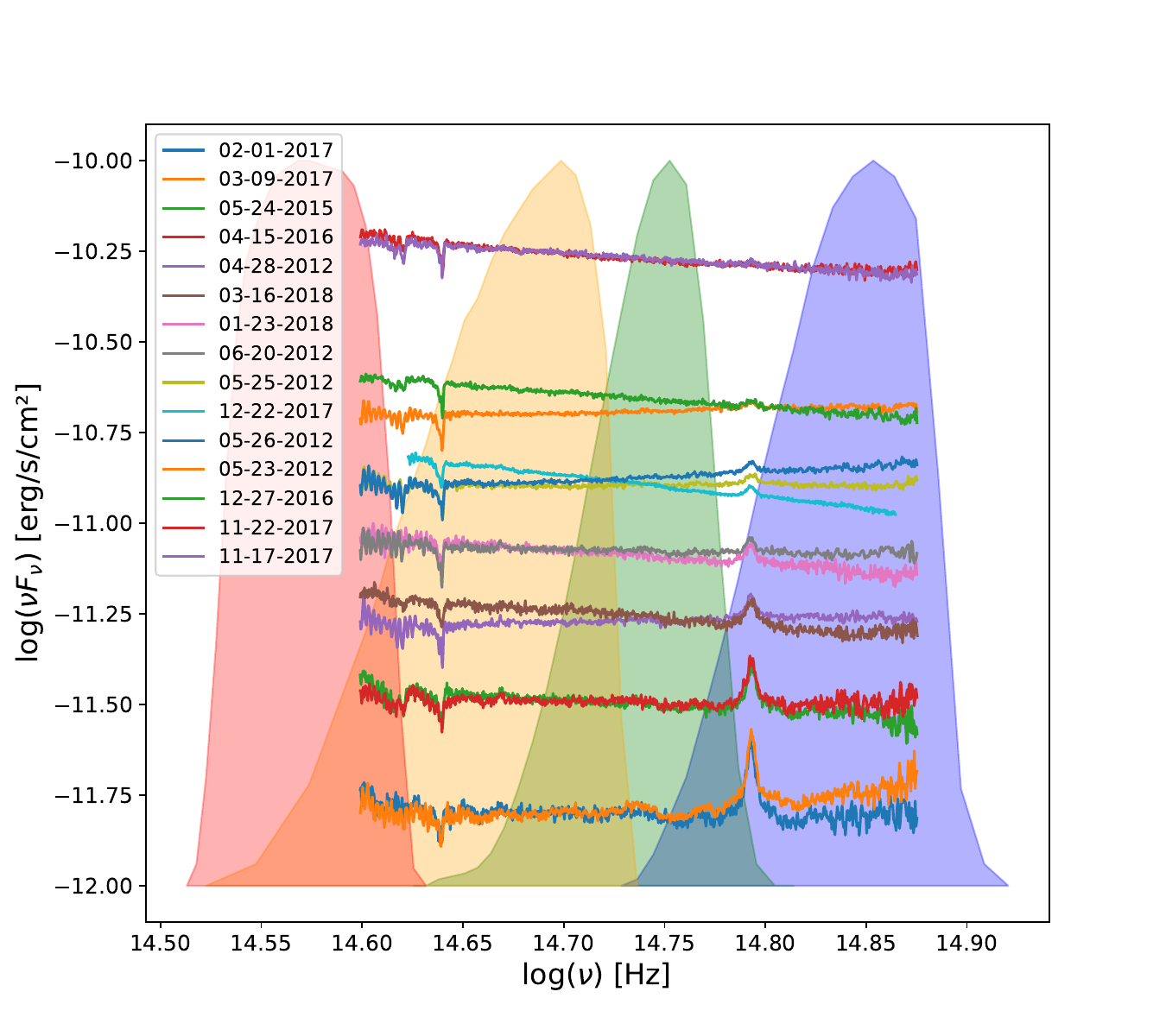}}
	\vspace{-3mm}
	\caption{Steward optical spectra of the blazar Ton 599 at different activity states.}
	\label{fig: Steward_spectra}
	\vspace{-1mm}
\end{figure}

\begin{figure}
    \centering
	\resizebox{10cm}{!}{\includegraphics{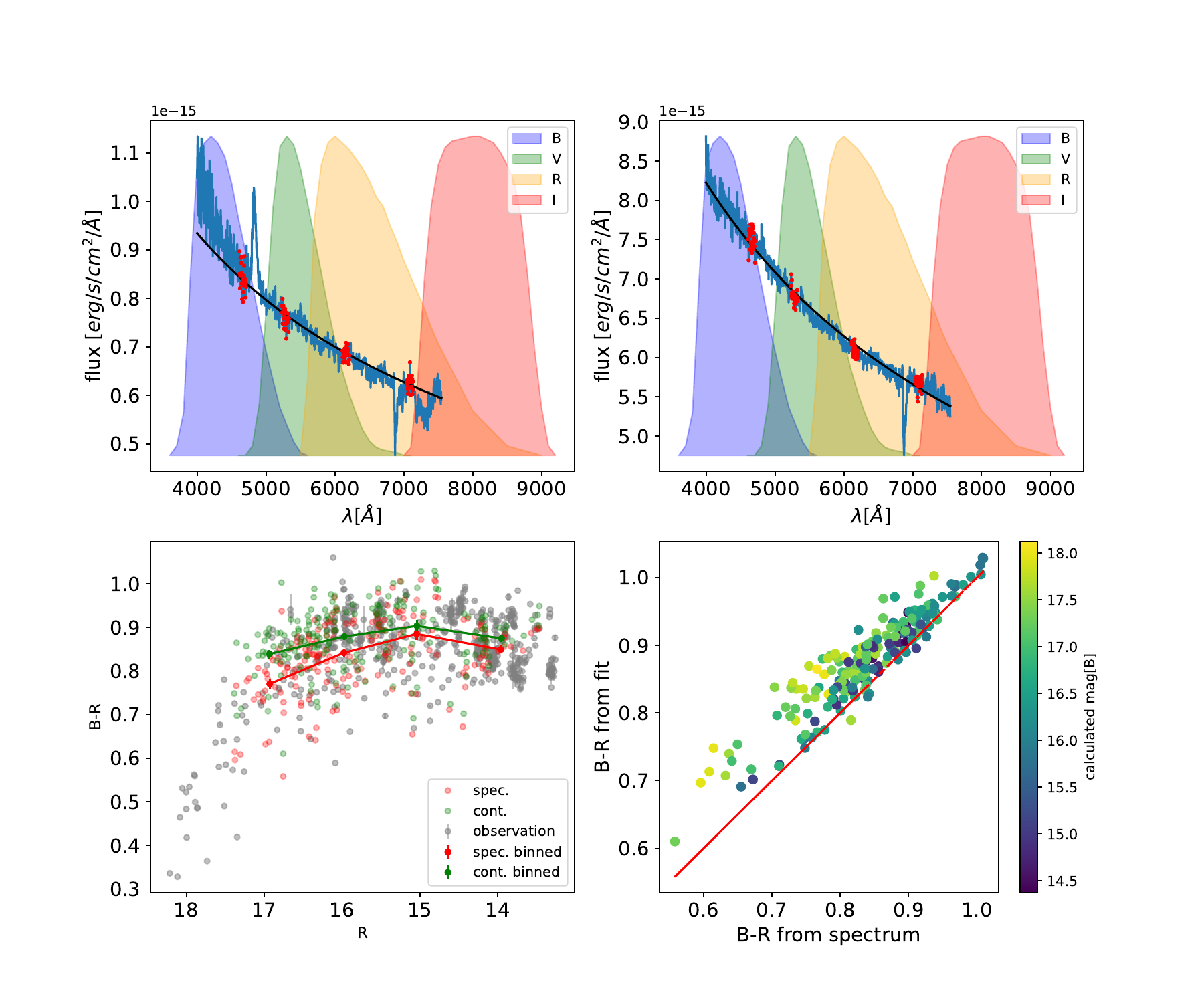}}
    \caption{Top: Steward spectra in low (left) and high (right) flux states, with continuum fits (black) and BVRI filter curves. Bottom left: (B–R) vs. R color–magnitude diagram from observed (gray) and model-based (red/green) data. Bottom right: comparison of observed and continuum-based color indices. The color bar encodes B magnitude.}
    \label{fig: BLR_vs_colors}
\end{figure}

\begin{figure}[htb]
\centering
\includegraphics[width=\columnwidth]{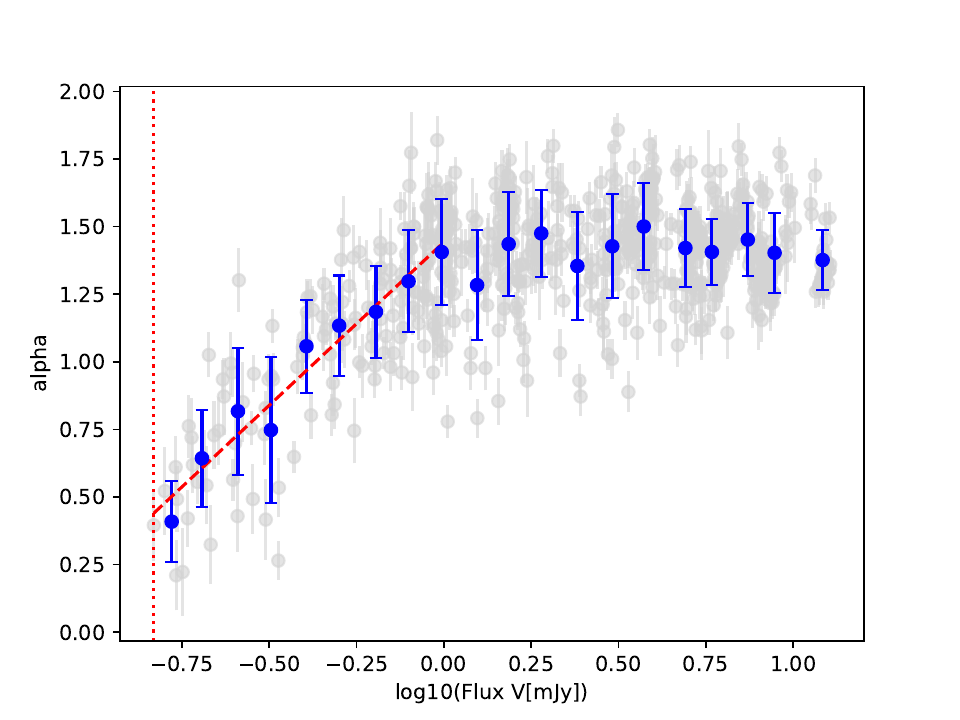}  
\caption{Binned spectral index $\alpha$ vs.\ $\log_{10}F_V$ for the $B\!V$ data.  Blue symbols are the bin averages with 1\,$\sigma$ errors; gray points are individual measurements.  The dashed red line shows the weighted linear fit for $\log_{10}F_V\le0$, and the vertical red line indicates $\log_{10}F_V$ at $\alpha_{\rm disk}=0.44$ (i.e.,\ $F_{\rm disk}=0.152\,$mJy).}
\label{fig:alphaBV}
\end{figure}

\indent Blazars show specific spectral characteristics. BL Lac blazars often show a bluer-when-brighter (BWB) long-term trend. Some authors found that long-term flux variations are quasi-achromatic and likely due to changes in the orientation of the jet, while the strongly chromatic short-term changes are attributed to energetic processes taking place in the optical jet (magnetic reconnection, shock)\citep[e.g.,][]{2021MNRAS.501.1100R, 2021MNRAS.504.5629R, 2023MNRAS.522..102R}.
\\
\indent Spectral behavior of the FSRQ type of blazars is somewhat different. In the case of CTA102, \citet{2017Natur.552..374R} found RWB trend in the faint state, switching to a mild BWB trend above R$\sim$15 mag. A similar behavior was described by \citet{2006A&A...453..817V} for the blazar 3C 454.3 and was interpreted as the influence of radiation from the disk in the faint state, which becomes less pronounced with the increase in blazar luminosity, where synchrotron radiation is dominant and the blazar exhibits BWB trend.
\\
\indent To study the spectral behavior of the blazar Ton 599, we used color indices and optical spectra. Color indices were calculated by combining data from the same observatory that were acquired within a time interval of 15 minutes. In the analysis we used only magnitudes with measurement uncertainties smaller than 0.03 mag. The number of color indices, which were obtained based on this criteria, ranges from $\sim$500 to $\sim$800 and is listed in the second column of Table \ref{tab: color_behaviour}. Magnitudes in these 15-min bins were calculated as weighted average of all magnitudes in the corresponding bins. For weights we used $w = 1 / magErr^{2}$, where magErr denotes the magnitude uncertainty of each data point.  Magnitude errors in bins were calculated according to the relation $\sqrt{1 /  \sum{magErr^{2}}}$. To obtain errors for the color indices, we added the magnitude errors in bins in quadrature.
\\
\indent Figure \ref{fig: color_VS_time_wLightcurve} shows the change in various color indices over time. The names of the color indices are shown in the legends of the subplots. For comparison, the first subplot from the top shows the light curve in the $R$-band. The red line is the best linear fit to the data obtained by the least-squares method. The results of the best linear fit, along with the fitting errors for each parameter, are given in Table \ref{tab: color_behaviour}. The table also lists the Pearson's linear correlation coefficient and the p value that tests the null hypothesis that the slope of the linear function is zero. These results suggest that there is no significant long-term color evolution for the blazar Ton 599.
\\ 
\\
\begin{table*}
\caption{Spectral analysis results of Ton~599 based on optical color indices}
\centering
\label{tab: color_behaviour}
\begin{tabular}{cccccccccc}
	\hline
	CI   &   N &   <CI>  &  <alpha> &        slope &   intercept &   slopeErr &   interceptErr &    r value &     p value \\
	\hline
	B-V  & 665 & 0.45 & 1.38 &  1.39e-05 &    0.40 & 2.53e-06 &      0.01 &  0.21 & 6.1e-08 \\
    B-R  & 547 & 0.84 & 1.30 &  2.64e-05 &    0.76 & 5.10e-06 &      0.02 &  0.22 & 3.2e-07 \\
	B-I  & 500 & 1.49 & 1.43 &  1.31e-05 &    1.45 & 6.10e-06 &      0.02 &  0.10 & 3.0e-02 \\
	V-R  & 815 & 0.40 & 1.21 &  9.51e-06 &    0.37 & 1.81e-06 &      0.01 &  0.18 & 1.6e-07 \\
	V-I  & 632 & 1.04 & 1.45 &  2.67e-06 &    1.04 & 3.47e-06 &      0.01 &  0.03 & 4.0e-01 \\
	R-I  & 635 & 0.65 & 1.64 & -9.82e-06 &    0.68 & 2.57e-06 &      0.01 & -0.15 & 1.0e-04 \\
	\hline
\end{tabular}
\tablefoot{CI is the color index, $N$ the number of combined data points, 
$\langle \mathrm{CI} \rangle$ the mean color index, and $\langle \alpha \rangle$ 
the average spectral slope. The remaining columns list the slope and intercept 
of the linear fit of color versus JD, their errors, the Pearson correlation 
coefficient ($r$), and the corresponding $p$ value.}
\end{table*}
\\
\indent The variation in the CI on smaller timescales is much more pronounced (see Figure \ref{fig: color_VS_time_wLightcurve}). Furthermore, it is striking that the amplitude of the color variation decreases with average flux density. To test this claim, we divided the data into 12 segments of about one year using the light curve in the $R$ band. As a measure of the color variation, we use the standard deviation of the colors in the segments and correlate this parameter with the corresponding mean flux density. The results of the correlation are shown in the Figure \ref{fig: dCI_VS_aveFlux} for B-V, B-R, and V-R color indices. The upper subplot shows the $R$-band light curve, where the shaded parts of the plot indicate the 12 segments in which we perform the correlation analysis. The red lines are the best linear fit to the data obtained by the least square method. The Pearson's correlation coefficients are -0.63, -0.59, and -0.58 for the B-V, B-R, and V-R color indices, respectively, as is indicated in the subplot legends. The p values, which tests the null hypothesis that the slope of the regression line is zero, are 0.03, 0.04, and 0.05 for the same order of the color indices. This result indicates that there is a significant negative correlation between the colors variations and the blazar brightness.
\\ 
\\
\indent Color-magnitude plots for various color indices are shown in the Figure \ref{fig: color_VS_mag}. The dashed blue lines on the subplots indicate the mean values of the color indices and they are 0.45, 0.84, 1.49, 0.40, 1.04, and 0.65 for the B-V, B-R, B-I, V-R, V-I, and R-I colors, respectively. The right side of the y axis shows the spectral index, which was calculated from the equation \ref{eq: alpha_from_flux} 
\begin{equation}
\alpha = {{0.4 \; CI - K}\over{log(\nu_{1}/\nu_{2})}},
\label{eq: alpha_from_magnitude}
\end{equation}
where subscripts 1 and 2 correspond to filter bands, K is a constant calculated from zero flux densities $ZP_{1}$ and $ZP_{1}$ according to relation $K = log(ZF_{1} / ZF_{2})$, and CI is the appropriate color index. The mean values of spectral slopes of the B-V, B-R, B-I, V-R, V-I, and R-I color indices are 1.38, 1.30, 1.43, 1.21, 1.45, and 1.64, respectively. The mean values of the color indices and $\alpha$ are summarized in the third and fourth columns of Table \ref{tab: color_behaviour}
\\
\\
\indent Figure \ref{fig: color_VS_mag} shows the dependence of the color indices and spectral slopes on 
magnitude. As can be seen, in 12 years of observation, the magnitude of the blazar Ton 599 has changed by about 5 mag in all filters. In the phase of weak blazar brightness, we can see a RWB trend for all color indices except for R-I. For these color indices, RWB switches to an achromatic trend above a certain brightness. On the other hand, color indices that include the I filter (B-I, V-I, and R-I), show transition from achromatic to BWB trend above a certain magnitude level. In order to quantitatively describe these trends, we fit linear curves to certain segments of the color-magnitude plots showing RWB, achromatic and BWB trends. We determined the boundaries that describe the transition from one trend to another visually. The results of the linear fitting are presented in the Table \ref{tab: color-magnitude_trends}. The table is divided into three groups corresponding to the RWB, achromatic, and BWB trends, which can be clearly discerned from the Figure \ref{fig: color_VS_mag}. For each of these groups, we have three columns that describe the Pearson correlation coefficient, the p value that tests the null hypothesis that there is no linear relationship between the two parameters we are correlating (p<0.05 rejects the null hypothesis), and the magnitude at which one trend changes to another. The results shown in the table suggest that the RWB, achromatic, and BWB trend are statistically significant.
\\
\begin{table*}
\centering
\caption{Color–magnitude correlation trends in Ton~599}

\label{tab: color-magnitude_trends}
\begin{tabular}{|c|c|c|c|c|c|c|c|c|c|}
\hline
& \multicolumn{3}{c|}{RWB trend} & \multicolumn{3}{c|}{Achromatic trend} & \multicolumn{3}{c|}{BWB trend}\\
& r & p & mag & r & p & mag & r & p & mag \\
\hline
B-V    & -0.74 & 3.0e-20 & B=17.1      & -0.03 & 4.8e-01 &    /       &  /   &   /      &/\\
B-R    & -0.84 & 2.4e-19 & R=16.4      &  0.14 & 2.0e-03 &    /       &  /   &   /      &/\\
B-I    & -0.57 & 3.4e-06 & B=17.1      &  0.03 & 6.9e-01 & B=15.2     & 0.55 & 1.6e-17  &/\\
V-R    & -0.70 & 3.8e-13 & R=16.6      &  0.13 & 3.0e-04 &    /       &  /   &   /      &/\\
V-I    & -0.57 & 2.0e-08 & V=16.7      &  0.08 & 1.3e-01 & V=14.8     & 0.60 & 5.0e-22  &/\\
R-I    &   /   &    /    &   /         &  0.08 & 1.4e-01 & R=14.6     & 0.46 & 1.6e-17  &/\\
\hline
\end{tabular}
\tablefoot{The table is divided into three sections: RWB, achromatic, and BWB trends 
(as shown in Figure \ref{fig: color_VS_mag}). In each case, $r$ is the Pearson correlation coefficient, 
$p$ is the probability of no correlation (the null hypothesis is rejected when $p<0.05$), 
and “mag” denotes the magnitude at which the transition between different trends occurs.}

\end{table*}
\\
\indent In order to better understand the color-magnitude trends, we show several spectra of the blazar Ton 599 at different flux levels in the Figure \ref{fig: Steward_spectra}. The spectra were taken from the publicly available archive of the Steward Observatory\footnote{\url{http://james.as.arizona.edu/~psmith/Fermi/}}, which were acquired as part of a monitoring program designed to support gamma-ray observations with the Fermi Gamma-ray Space Telescope \citep{2009arXiv0912.3621S}. Unfortunately, the Steward spectra were recorded from 2011 to 2018, so they do not cover Period III in which the most powerful outbursts occurred, but they cover the Period II well. For illustration, we selected representative spectra at different flux levels. In the observation frame, the spectra range from $\sim$4000 $\AA$ to $\sim$7600 $\AA$, which corresponds to $\sim$2300 $\AA$ - $\sim$4400 $\AA$ in the rest frame of the blazar. Therefore, the broad emission line, which is pronounced in the low activity phase, is Mg II $\lambda2798$ originating from the broad line region (BLR) of the AGN. In addition to the strong Mg II emission line, broad Fe-complex features are clearly visible, which also originate from the AGN \citep[e.g.,][]{2022ApJ...926..180H}. Absorption lines in the infrared part of the spectrum are atmospheric lines\footnote{\url{http://james.as.arizona.edu/~psmith/Fermi/DATA/READMEspec.html}}. In addition to the spectra, the figure shows the transmission curves of the Johnson-Cousin filters that were used for photometric observations and are marked with transparent blue, green, orange, and red colors for the B, V, R, and I filters, respectively.
\\
\\
\indent From the spectra shown in Figure \ref{fig: Steward_spectra}, several things can be noted. The first is that the contribution of thermal radiation is significant in the phases of weak blazar brightness. As the mean flux increases, the thermal component weakens and synchrotron radiation becomes increasingly dominant. This spectral behavior can explain the RWB trend we observe in the color-magnitude plots. Then, we see a diverse behavior of the spectra for (approximately) the same mean flux level, which explains the large dispersion of the color indices for a certain magnitudes that can be seen in the color-magnitude plots. Colors involving the I band show a large dispersion and this makes the BWB trend suspect.
\\
\indent In order to test if the thermal radiation may be responsible for the RWB trend of the color-magnitude diagram, we use Steward spectra. In the first step, we fit the spectra with a power law function $F=F_{o}(\lambda / \lambda_{o})^{\alpha}$ through the points representing the continuum, where $F_{o}$ is the normalization parameter at $\lambda_{o} = 3000 \AA$ in the rest frame and $\alpha$ is the spectral index. Both $F_{o}$ and $\alpha$ are free parameters. The obtained best fit represents the spectrum without Mg II and Fe emission lines; that is, the continuum. Then we do convolution of both spectra, the continuum and the original, with the BVRI filter transparency curves to determine the corresponding magnitudes. The top two panels of the Fig. \ref{fig: BLR_vs_colors} show two spectra from the Steward archive, one corresponding to the low flux state of the blazar (left), and the other to a relatively high flux state (right). Data points we used for fitting the power law function, i.e., the continuum windows, are marked in red colors. The two middle continuum windows (5226 - 5329 \AA, 6105 - 6209 \AA in the observed frame) are taken from \citet{2022ApJ...926..180H}. The remaining two continuum windows were not taken from the literature, but were defined with the aim of eliminating the emission and absorption lines in the B and I filters as best as possible. We have also marked the BVRI transparency curves with the colors indicated in the plot legends. The black line represents the best fit power law functions for these two spectra. The lower left panel shows the B-R color versus R magnitude plot. Observations are marked with transparent gray circles. Transparent red and green circles represent calculated colors/magnitudes; red ones correspond to filter convolution with the original Steward spectra, and green ones correspond to convolution with the continuum. For better visualization, with red and green lines we present the binned values of colors and magnitudes with bin widths of 1 magnitude. The lower right panel compares the calculated color indices. The diagonal red line is the equality line y=x. The points are color-coded so that one can see how color indices depend on the brightness of the blazar. Both lower panels indicate that the color indices tend to be redder when the contribution from the thermal component—likely associated with the accretion disk emission in the optical–UV range—is not included, and that this trend becomes more pronounced at fainter flux levels (i.e., higher magnitudes).
\\
\\
\indent \citet{2001AJ....122..549V} built a composite quasar spectrum from over 2200 SDSS spectra (rest frame 800–8555$\AA$) and showed that, between $\sim1300\;\AA$ and 5000 $\AA$, the continuum follows $F_\nu\propto\nu^{-0.44}$.  We adopt this as the “disk” index $\alpha_{\rm disk}=0.44$. \citet{2013ApJ...773..147J} analyzed the blazar 3C 454.3 and measured the continuum spectral index $\alpha_{\rm cont}$ as a function of the $V$-band flux $S_V$ using simultaneous photometry (within $\sim$1 h). They showed that the blue (disk) component remains constant while the red (jet) component varies, fit a linear relation over the range $0 \le S_V \le 4,$mJy, and inverted it at $\alpha_{\rm disk} = 0.44$ to estimate the disk flux as $S_{\rm disk} \simeq 0.85 \pm 0.15$ mJy in V (and $0.91 \pm 0.16,$mJy in R). We followed a similar procedure on our two-filter photometry using the BV, BR, and BI pairs. The magnitudes were first converted to flux densities using standard zero-flux values. Errors were propagated using the relation $\sigma_F = 0.4 \ln(10) F \sigma_m$. For each filter pair, we then computed the spectral index as $\alpha = -\ln(F_1/F_2)/\ln(\nu_1/\nu_2)$. The resulting $\alpha$ values were binned in $\log_{10}F_2$, and we performed a weighted linear fit restricted to $\log_{10}F_2 \le 0$. Finally, we solved for $F_2$ at a fixed value of $\alpha_{\rm disk} = 0.44$.  As an example, Figure~\ref{fig:alphaBV} shows the binned spectral index $\alpha$ versus $\log_{10}F_V$ for the BV pair.  Blue symbols are the bin means with 1\,$\sigma$ errors; gray points are individual measurements.  The dashed red line is the weighted linear fit for $\log_{10}F_V\le0$, and the vertical red line marks the derived $\log_{10}F_V$ at $\alpha_{\rm disk}=0.44$ (i.e., $F_{\rm disk}=0.152\,$mJy).  Table \ref{tab:disk_flux_Bx} summarizes the results for the three $B$-anchored pairs.
\\
\begin{table}[htb]
\caption{Fit parameters and derived disk fluxes for $\alpha_{\rm disk}=0.44$.}
\label{tab:disk_flux_Bx}
\centering
\begin{tabular}{l  c  c  c}
\hline\hline
Filter & $m\pm\sigma_m$ & $b\pm\sigma_b$ & $F_{\rm disk}\pm\sigma_F$ [mJy] \\
\hline
BV & $1.239\pm0.235$ & $1.454\pm0.114$ & $0.152\pm0.063$ \\
BR & $1.427\pm0.258$ & $1.448\pm0.094$ & $0.197\pm0.065$ \\
BI & $1.243\pm0.246$ & $1.545\pm0.086$ & $0.129\pm0.056$ \\
\hline
\end{tabular}
\end{table}
\\
\\
\noindent All three estimates of $F_{\rm disk}$ (0.13–0.20 mJy) lie well below the minimum observed flux ($\gtrsim1\,$mJy), confirming that the thermal disk (big blue bump) contributes only a few tenths of a millijansky, while the variable nonthermal jet dominates at higher brightness.  The slight increase in $F_{\rm disk}$ from BV to BR is consistent with a redder jet spectrum varying on top of a nearly constant blue disk component, in agreement with the two-component model of \citet{2013ApJ...773..147J} and \citet{2007A&A...473..819R}.

\subsection{Estimation of the black hole mass}

\begin{figure}
    \centering
	\includegraphics[height=6cm,width=\columnwidth]{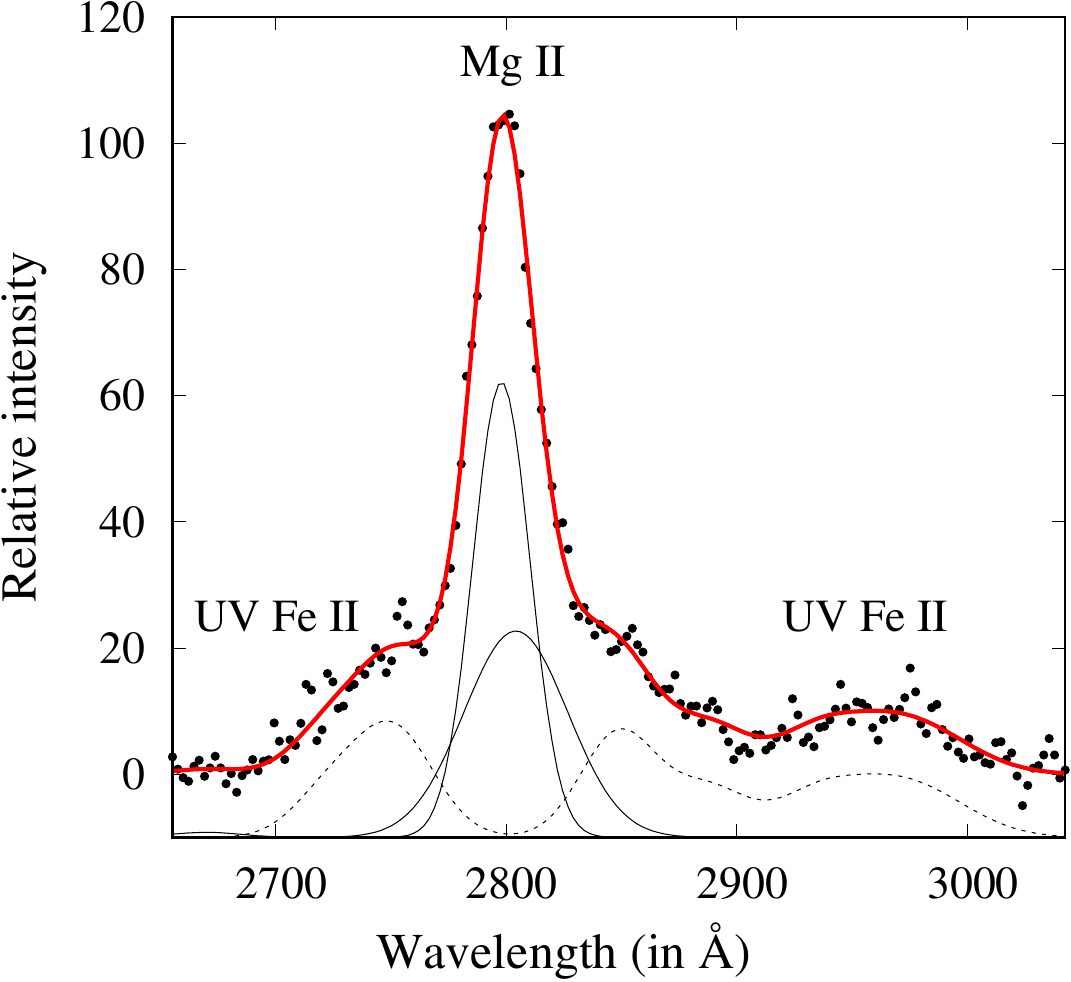}
    \vspace{-3mm}
    \caption{Example of the Mg II + UV Fe II fit in the 2650–3050 \AA\ range for the Steward spectrum observed on February 7, 2011. The core and wing Gaussian components of the Mg II line are plotted with solid lines, while the UV Fe II template from \cite{2019MNRAS.484.3180P} is plotted with a dashed line. The observed data are represented by dots, and the red line indicates the best fit.}
    \label{fig: UV_fit_example}
    \vspace{-1mm}
\end{figure}

\indent We determined the black hole mass of Ton 599 using the Steward spectra and the empirical relation that connects the black hole mass with the measured parameters of the Mg II emission line \citep{2006ChJAA...6..396K}:
\[
\frac{M_{BH}}{M_\odot} = 2.9 \times 10^6  \left(\frac{L_{MgII}}{10^{42} \text{ erg/s}}\right)^{0.57 \pm 0.12} \left(\frac{FWHM_{MgII}}{10^3  \text{ km/s}}\right)^{2},
\]
where \( L_{MgII} \) is the line luminosity and \( FWHM_{MgII} \) is the full width at half maximum (FWHM) of the line.
\\
\indent To accurately measure the Mg II spectral parameters needed for M$_{BH}$ calculation, it is of great importance to extract the profile of the Mg II line, which overlaps with numerous UV Fe II lines. The additional complexity of the Mg II line is that it is actually a doublet ($\lambda 2796 \AA$  and $\lambda 2803 \AA$) with an intensity ratio dependent on the optical depth of the emitting region. Given that the doublet separation ($\sim 8 \AA$) is much smaller than the total line width, Mg II can be treated as a single line, as suggested by \cite{2013A&A...555A..89M}.
We performed a multicomponent fit in the 2650 - 3050 \AA \ spectral range, as described in detail in \cite{2019MNRAS.484.3180P}. After subtracting the continuum, the Mg II emission line was fitted with a two-Gaussian model, where one Gaussian fits the core and one the wings of the Mg II line. In this way, we were able to fit the asymmetry well in the Mg II line. The UV Fe II lines in 2650-3050 \AA \ range were fit with the complex UV Fe II template described in \cite{2019MNRAS.484.3180P}. In that model, the UV Fe II lines are divided into six groups of Fe II lines, which have different intensity parameters, while the widths and the shifts of all UV Fe II lines are taken to be the same. The total flux of the Mg II line, as well as the FWHM(Mg II) used for the M$_{BH}$ estimation, were obtained using the total Mg II profile. The flux was calculated as the sum of the two Gaussian components from the best fit. The FWHM of the total Mg II profile was measured by normalizing the peak of the summed two-Gaussian fit to unity and determining its half-maximum width. Figure \ref{fig: UV_fit_example} shows an example of the Mg II + UV Fe II fit in the 2650–3050 \AA\ range for the Steward spectrum observed on February 7, 2011.
\\
\\
\indent To determine the black hole mass, we selected ten spectra with the lowest continuum flux, minimizing the jet contribution while maximizing the contribution from the broad-line region (BLR). The measured parameters are listed in Table \ref{tab:M_BH}, with associated standard errors from the fit. The uncertainties in the individual black hole mass estimates were obtained by propagating the measurement uncertainties through the calibration relation. For the final black hole mass and its uncertainty, we adopted the mean value from Table \ref{tab:M_BH} and the mean propagated uncertainty, yielding $\log(M/M_{\odot}) = 8.77 \pm 0.26$. The obtained value is consistent with typical SMBH masses of $10^6-10^9 M_{\odot}$, exceeding the $10^8 M_{\odot}$ typical of radio-loud AGNs \citep{2011MNRAS.416..917C}. It is also comparable to the result by \cite{2019PhDT.......127K}, which for the black hole mass of Ton 599 obtained $\sim9\times10^{8} M_{\odot}$. The Eddington luminosity is given by \citep[e.g.,]{1985PhR...123..117W}
\[
L_E = 1.3 \times 10^{38} \left({\frac{M_{BH}}{M_{\odot}}}\right) \;\; [erg/s],
\]
which for our black hole mass estimate gives $\sim 7.7 \times 10^{46}$ erg/s.
\\
\begin{table*}[h!]
\caption{\footnotesize Mg\,II line measurements and black hole mass estimates for Ton~599}
\centering
\label{tab:M_BH}
\begin{tabular}{lllll}
	\toprule
	Date & $F_{\text{cont}} \pm \text{err}$ & $F_{\text{MgII}} \pm \text{err}$ & $\text{FWHM} \pm \text{err}$ & $\log(M_{BH}) \pm \text{err}$ \\
	\midrule
	2011-01-04 &               $199.98 \pm 16.18$ &             $4522.58 \pm 659.72$ &         $3632.78 \pm 126.91$ &               $8.74 \pm 0.25$ \\
	2011-02-02 &               $206.84 \pm 16.46$ &             $4860.20 \pm 677.99$ &         $3766.23 \pm 126.41$ &               $8.79 \pm 0.25$ \\
	2015-05-25 &               $216.29 \pm 24.11$ &            $5769.69 \pm 1455.34$ &         $3918.44 \pm 236.83$ &               $8.87 \pm 0.27$ \\
	2011-03-04 &               $218.37 \pm 14.40$ &             $4383.97 \pm 388.69$ &         $3758.71 \pm 183.37$ &               $8.76 \pm 0.25$ \\
	2018-06-14 &               $225.77 \pm 43.57$ &            $3766.39 \pm 1530.84$ &         $3407.78 \pm 227.34$ &               $8.64 \pm 0.26$ \\
	2015-06-21 &               $250.77 \pm 50.69$ &            $4918.12 \pm 2192.63$ &         $3473.29 \pm 354.25$ &               $8.72 \pm 0.29$ \\
	2011-02-07 &               $253.16 \pm 16.56$ &             $4748.14 \pm 704.07$ &         $3727.63 \pm 131.91$ &               $8.77 \pm 0.25$ \\
	2016-05-08 &               $258.25 \pm 34.26$ &            $4679.48 \pm 2178.96$ &         $3650.43 \pm 296.79$ &               $8.75 \pm 0.28$ \\
	2015-05-20 &               $266.19 \pm 49.11$ &            $5177.43 \pm 1556.71$ &         $3906.15 \pm 270.31$ &               $8.84 \pm 0.27$ \\
	2016-04-14 &               $266.39 \pm 34.28$ &            $5443.40 \pm 1554.16$ &         $3666.73 \pm 255.32$ &               $8.79 \pm 0.27$ \\
	\bottomrule
\end{tabular}
\tablefoot{Columns list the observation date, the continuum flux density at 
3020\,\AA\ ($F_{\mathrm{cont}}$, in units of $10^{-17}$ erg s$^{-1}$ cm$^{-2}$ \AA$^{-1}$), 
the total Mg\,II emission line flux ($F_{\mathrm{MgII}}$, in units of 
$10^{-17}$ erg s$^{-1}$ cm$^{-2}$), the FWHM of the Mg\,II line 
(in km\,s$^{-1}$), and the logarithm of the black hole mass ($\log M_{\mathrm{BH}}$) 
in solar masses. Uncertainties are standard errors of the measurements.}
\end{table*}

\section{Discussion}

\paragraph{Flux distribution and RMS-flux relation.} The statistical properties of long-term optical variability in blazars, such as the flux distribution and the RMS--flux relationship, offer profound insights into the underlying emission mechanisms and the physical processes at play within their relativistic jets. Our comprehensive analysis of Ton 599 over a 12-year period reveals characteristics that consistently point toward multiplicative processes governing its variability on these extended timescales.
\\
\indent The flux distribution of Ton 599, when examined over the entire observational period, is best described by a log-normal distribution across all BVRI bands, rather than a normal (Gaussian) distribution. This finding is a strong indicator of multiplicative processes, where variations are compounded. Such log-normal behavior in long-term optical blazar light curves is corroborated by studies such as \cite{2021ApJ...923....7B}, who analyzed a sample of 12 $\gamma$-ray-bright blazars with decade-long optical data and similarly found their flux distributions to be well-characterized by log-normal probability density functions (PDFs). This was also interpreted as evidence of multiplicative nonlinear processes driving the variability. The log-normal nature of blazar flux has also been noted in other bands, such as $\gamma$-rays \citep[e.g.,][]{2018RAA....18..141S, 2020ApJ...891..120B}. 
\\
\indent Complementing the flux distribution analysis, the RMS-flux relationship in Ton 599 shows a significant positive linear trend in all optical bands (see Table \ref{tab: RMS-flux relation} and Figure \ref{fig: fluxPDF_fluxRMS_relations}). This linear scaling, where the amplitude of variability increases with the mean flux, provides further robust evidence for dominant multiplicative processes. Again, this aligns with the findings of \cite{2021ApJ...923....7B} for their sample of blazars, where linear RMS-flux relations were observed in the optical domain, supporting the multiplicative scenario. The combination of a log-normal flux distribution and a linear RMS-flux relation is widely considered a strong signature of multiplicative processes in accreting astrophysical systems \citep[e.g.,][]{2005MNRAS.359..345U}. Such behavior is thought to arise from various physical scenarios. \cite{2021ApJ...923....7B} discusses several such mechanisms, including the general concept of multiplicative coupling of perturbations occurring within the accretion disk and/or the jet. More specifically, models invoking propagating fluctuations within the accretion disk, where uncorrelated variations in viscosity (the $\alpha$-parameter) at different radii modulate the accretion rate as they propagate outward \citep{1997MNRAS.292..679L}, can account for these observations. These disk-borne modulations can subsequently propagate into the blazar jet via a strong disk-jet connection  and are then significantly amplified by relativistic beaming and projection effects. Alternatively, processes within the relativistic jet itself, such as the “jets-in-jets” scenario whereby Poynting-flux-dominated jets might generate numerous isotropically distributed mini-jets \citep{2009MNRAS.395L..29G}, have also been proposed. Such mini-jet emission has been shown to produce highly skewed flux distributions and adhere to the RMS-flux relation. Our findings for Ton 599 are consistent with these general frameworks proposed for blazar variability. 
\\
\indent It is noteworthy that a single, geometric interpretation can provide a compelling explanation for both of these findings. As has been proposed by \citet{2024A&A...692A..48R}, changes in the viewing angle to the jet naturally lead to the observed effects; a decrease in the viewing angle enhances the Doppler beaming, which not only increases the observed flux, contributing to the log-normal distribution, but also amplifies the intrinsic variability, thus producing the linear relationship between RMS and flux that we observe.
\\
\indent In our detailed examination of the RMS-flux relation for Ton 599, the intercept of the linear fit was treated as a free parameter, and its value exhibited some dependence on the specific filter and binning method employed (Table \ref{tab: RMS-flux relation} in this work). Using the “size correlation” method, statistically significant positive intercepts on the RMS axis were observed for the V ($0.21 \pm 0.09$), R ($0.17 \pm 0.07$), and I ($0.54 \pm 0.21$) bands. While a literal interpretation of nonzero RMS at zero flux is generally considered nonphysical, such positive intercepts can, as discussed in studies such as \cite{2021ApJ...923....7B}  (who references \cite{2004A&A...414.1091G}), indicate the presence of an additional variability component that may become more prominent at very low flux states. Alternatively, it may suggest that the true relationship deviates from perfect linearity at the lowest flux levels probed. The B-band intercept ($0.09 \pm 0.11$) with this binning method was consistent with zero. In contrast, using fixed 50-day time intervals for binning (“time correlation”), the intercepts for the B ($-0.02 \pm 0.06$), V ($0.06 \pm 0.08$), and R ($0.00 \pm 0.06$) bands were consistent with zero, supporting a more direct proportionality between RMS and flux in these cases. The I band showed a marginally positive intercept ($0.20 \pm 0.14$) with this latter method. These nuances in the intercept values might reflect the sensitivity of this parameter to the binning approach and the consequent sampling of low-flux states. However, the robustly determined positive slope of the RMS-flux relation across all filters and both binning methods consistently points to the multiplicative nature of the dominant LTV in Ton 599.
\\
\indent The characterization of LTV in Ton 599 contrasts with its behavior on shorter timescales. Our analysis of intranight light curves indicates that these rapid (hourly) fluctuations are well represented by a Gaussian distribution, suggesting that additive processes are likely dominant on these very short timescales. This finding is in agreement with other studies, such as \cite{2024A&A...686A.228O}, who reported Gaussian flux distributions for IDV in TESS observations of blazars, linking them to linear (additive) processes in compact jet regions.
\\
\indent Bridging these extremes, the study by \cite{2023MNRAS.518.1459P} using $\sim$25-day continuous TESS light curves provides valuable context for intermediate timescales. While their general finding for a sample of 29 blazars was that optical flux histograms on these day-to-week timescales are often consistent with normal (Gaussian) PDFs (frequently bimodal), their specific result for Ton 599 (Sector 22) indicated that a log-normal distribution provided a better fit than a normal one, based on AIC and BIC criteria (see Table 3 in \cite{2023MNRAS.518.1459P}). This persistence of log-normality for Ton 599, from $\sim$25-day segments \citep{2023MNRAS.518.1459P} up to the multiyear periods analyzed in this work, is particularly compelling.
\\
\indent Thus, a hierarchical picture of variability emerges for Ton 599: additive processes appear to govern the most rapid, hourly fluctuations (IDV), while multiplicative processes become dominant as the timescale extends from weeks \citep[][for Ton 599]{2023MNRAS.518.1459P} to years (this work). This makes Ton 599 an interesting object where the signature of multiplicative processes is robustly observed even on relatively short, continuous observational segments, setting it apart from the average behavior seen in the TESS sample by \cite{2023MNRAS.518.1459P}  and highlighting the persistent nature of these mechanisms in this particular blazar. This complex, timescale-dependent nature of flux statistics underscores the interplay of different emission mechanisms or conditions within blazar jets.
\\
\indent It is important to note, however, that \cite{2020ApJ...895...90S} has cautioned that certain classes of stationary, linear, additive time-series models can also, under specific conditions (e.g., with a particular choice for the distribution of innovations), reproduce statistical properties such as log-normal flux distributions and linear or near-linear RMS-flux relationships. This suggests that while the combination of these properties observed in Ton 599 strongly points toward a multiplicative origin in line with prevailing interpretations \citep[e.g.,][]{2005MNRAS.359..345U, 2020ApJ...891..120B}, definitive conclusions about the underlying physical processes (i.e., ruling out all possible linear, additive scenarios) solely based on these flux statistics should be made with awareness of these potential mathematical degeneracies. Therefore, the interpretation of multiplicative processes is further strengthened when considered alongside other observational evidence from blazars, such as spectral variability, polarization characteristics, and higher-order timing statistics, which may provide more direct constraints on the nonlinear and coupled nature of the emission mechanisms.

\paragraph{Power spectral density.} Our PSD analysis of the long-term optical light curves of Ton 599 reveals that its variability is characteristic of a red-noise process across all BVRI bands. The PSDs are described well by a single power-law model, $P(\nu) \propto \nu^{-\beta}$, with derived slopes $\beta$ ranging from approximately 1.40-1.79 (e.g., $\beta_{R}$ = 1.40±0.05). Such slopes, typically falling in the range $1 \le \beta \le 3$ for blazar optical variability, indicate that the flux variations are strongly correlated over time, with more power concentrated on longer timescales. This is a hallmark of stochastic processes where past variations influence future ones.
\\
\indent The specific slopes we find for Ton 599 are broadly consistent with those reported for other blazars in the optical domain based on long-term monitoring. For instance, our R-band slope of $\beta_{R} \approx 1.40$ is remarkably similar to the average optical PSD slope of $\beta_{R} \approx 1.42$ reported by \cite{2018A&A...620A.185N} for a large sample of blazars. Other studies of long-term optical variability have also found comparable slopes; for example, \cite{2017ApJ...837..127G} reported a long-term optical (R-band) PSD slope of $\beta = 2.0 \pm 0.1$ for PKS 0735+178 spanning 23 years. Earlier work on PKS 0735+178 by \cite{2007A&A...467..465C}, using structure function analysis, also found PSD slopes between 1.5 and 2.0 for long-term optical data. Analyses of Kepler optical data for blazars yielded $\beta \sim 1.5-2.0$ \citep{2013ApJ...766...16E, 2014ApJ...785...60R}. The PSDs in this work were derived by fitting the periodogram of linearly interpolated light curves, applying mean subtraction and Hanning windowing to minimize leakage, a methodological approach also adopted in other optical blazar studies confronted with unevenly sampled data \citep[e.g.,][]{2017ApJ...837..127G}.
\\
\indent A key result from our PSD analysis is the absence of any statistically significant peaks indicative of quasi-periodic oscillations (QPOs), or distinct breaks signifying characteristic timescales, within the frequency range probed by our decade-long dataset. This lack of pronounced features reinforces the interpretation that the variability of Ton 599 is predominantly stochastic. This finding is consistent with many blazar studies; for instance, \cite{2023MNRAS.518.1459P} also reported no significant PSD breaks or QPOs in their analysis of short ($\sim$25-day) TESS light curve segments for their blazar sample. Similarly, \cite{2017ApJ...837..127G} did not detect any low-frequency flattening or high-frequency cutoffs in their composite optical PSD for PKS 0735+178, which extended down to minute timescales. While some blazars do exhibit QPOs in their optical light curves on year-like timescales \citep[e.g.,][]{2021ApJ...923....7B} for sources such as S5 0716+714 and Mrk 421), Ton 599 does not show such persistent periodic features in our current, extensive dataset. We also note that earlier work by \cite{2006PASJ...58..797F} suggested potential periodicities of 1.58 and 3.55 years for Ton 599 from R-band data spanning 1974-2002. Our more recent and extensive dataset does not provide evidence supporting these specific periodicities.
\\
\indent The derived PSD slopes for Ton 599 can also be considered in the context of its blazar classification and spectral characteristics. Ton 599 is a flat spectrum radio quasar (FSRQ), a class of blazar whose synchrotron peak frequency ($\nu_{peak}^{syn}$) typically lies in the low-synchrotron peaked (LSP) range (i.e., $< 10^{14}$ Hz). While this provides a general classification, recent detailed multiwavelength spectral energy distribution (SED) modeling of Ton 599 by \cite{2024MNRAS.529.1356M} during its exceptionally bright flare in January 2023 has revealed a more dynamic picture of its synchrotron peak. Analysis of their fit synchrotron components across different flux states during this flare \citep[visually inferred from Figure 5 in][]{2024MNRAS.529.1356M} suggests that $\nu_{peak}^{syn}$ can shift to higher frequencies, reaching or even exceeding $10^{14}$ Hz, particularly during periods of heightened activity. This indicates that Ton 599 can transition toward or into the intermediate-synchrotron peaked (ISP) regime during such active phases, highlighting the dynamic nature of its SED. Our long-term optical PSD slopes (e.g., $\beta_R \approx 1.40$) are derived from data spanning over a decade, thus representing an average characteristic of Ton 599's variability across its diverse activity states, which would encompass both quiescent periods where it may behave as a typical LSP, and more active phases where it might exhibit ISP characteristics. 
\\
\indent This observed dynamic behavior of $\nu^{syn}_{peak}$ in Ton 599 is relevant when considering studies such as that by \cite{2018A&A...620A.185N}, who investigated the relationship between long-term optical PSD slopes ($\beta$) and $\nu^{syn}_{peak}$ for their blazar sample and found no significant correlation. The PSD characteristics of Ton 599, an FSRQ exhibiting such spectral peak variability, are therefore consistent with this general finding from \cite{2018A&A...620A.185N} of a lack of a simple, direct dependence of the long-term optical PSD slope on a single, fixed synchrotron peak frequency. Instead, the PSD slope likely reflects more fundamental properties of the energy dissipation and particle acceleration processes that persist across these varying spectral states.
\\
\indent A comparison of our optical PSD slopes with those obtained at radio and $\gamma$-ray energies provides additional context. 
In the radio domain, \citet{Kankkunen2025} analyzed a multi-decade 15 GHz OVRO light curve of Ton~599 and reported a clear break in the PSD at $\sim$1500 days, interpreted as a transition from a steep red-noise process ($\beta \approx 2$) on shorter timescales to a flatter slope ($\beta \approx 1$) on longer timescales. 
In this work, based on $\sim$4300 days of optical monitoring, we find that the BVRI light curves are well described by single power-law PSDs with slopes $\beta_{\rm opt} \approx 1.4$--$1.8$, without statistically significant evidence for a break. 
Although hints of a steepening above $\sim$1000 days may be present, the limited baseline prevents us from confirming a robust feature comparable to the radio break. At $\gamma$-ray energies, \citet{Wang2024} performed a systematic Fermi-LAT analysis of TeV blazars including Ton~599, and derived PSD slopes for monthly binned light curves (their Table~6). 
For FSRQs the average slope is close to unity ($\beta_{\gamma} \approx 1.0$), consistent with flicker noise. 
This is flatter than both our optical slopes ($\beta_{\rm opt} \approx 1.4$--$1.8$) and the radio slope below the 1500-day break ($\beta_{\rm radio} \approx 2$), suggesting a wavelength-dependent steepening from $\gamma$-ray to optical to radio frequencies. The overall picture indicates that variability processes in Ton~599 become progressively more correlated (i.e., longer memory, steeper PSD) when moving from high to low energies. 
This trend may reflect different dominant timescales for particle acceleration and cooling in the jet: $\gamma$-ray variability is driven by fast processes in compact regions, optical variability (this work) arises from intermediate-scale synchrotron zones, while the radio emission integrates over larger jet volumes, where long-term jet-dynamics and propagation effects dominate.
\\
\indent Finally, while a power-law PSD provides a useful mathematical description of Ton 599's observed variability spectrum, it is important to acknowledge the cautionary notes raised in the literature. As has been pointed out by \cite{2018A&A...620A.185N} and more broadly by \cite{2020ApJ...895...90S}, this statistical descriptor does not unequivocally define the underlying physical process as simple power-law noise. The observed PSD could potentially arise from a superposition of discrete events or flares, or even from certain classes of linear, additive models under specific conditions. Thus, the interpretation of the PSD slopes in terms of specific physical mechanisms such as turbulence or propagating accretion instabilities is best supported when considered alongside other observational characteristics of Ton 599 discussed in this work.

\paragraph{IDV analysis.} Our analysis of IDV in Ton 599, based on observations where significant flux changes were detected on hourly timescales (Sect. 3.2 of this work), provides estimates for key physical parameters of the emitting regions. The characteristic variability timescales ($\tau_{\mathrm{var}}$) observed range from approximately 0.3 hours to 12.5 hours. Using these timescales and an adopted mean Doppler factor of $\delta$ = 23.1, we derived upper limits on the size of the emission region ($R$) and the strength of the magnetic field ($B$).
\\
\indent The IDV-derived emission region sizes (R) for Ton 599 range from $9.8 \times 10^{15}$ cm to $4.3 \times 10^{17}$ cm, with a mean value around $2 \times 10^{17}$ cm. These can be compared with estimates of the emission region size from broadband SED modeling. For instance, \cite{2018ApJS..235...39C} derived an emission region radius of $R \approx 7.9 \times 10^{16}$ cm for Ton 599. While \cite{2010MNRAS.402..497G} primarily list the dissipation distance from the central black hole ($R_{diss}$) for Ton 599 as $1.14 \times 10^{17}$ cm, their underlying model assumes that beyond an initial acceleration zone, the jet becomes conical with a semi-aperture angle $\psi \approx$ 0.1 radians. This allows for an estimation of the transverse size (radius) of the emission region ($r_{diss}$) in their model as $r_{diss} \approx \psi \times R_{diss} \approx 1.14 \times 10^{16}$ cm. Our IDV-derived region sizes, particularly at the lower end of the range, are thus remarkably comparable to this estimated size from the \cite{2010MNRAS.402..497G} model parameters, and the size derived by \cite{2018ApJS..235...39C} also falls within our IDV estimates. This broad agreement suggests that the rapid flares observed on intraday timescales might originate from compact regions whose dimensions are consistent with those implied by one-zone leptonic models. It is plausible that IDV probes these entire compact dissipation zones during coherent variations, or perhaps more localized, very dense subregions within a larger emitting volume, aligning with commonly inferred blazar emission region sizes of $10^{16} - 10^{17}$ cm \citep[e.g.,][]{1996MNRAS.280...67G, 1999A&A...352...19R}. The range of our derived emission region sizes, $R \approx 9.8 \times 10^{15} - 4.3 \times 10^{17}$\,cm can be further contextualized by a focused comparison with studies of the major November 2017 flare (p4) by \cite{2018ApJ...866..102P, 2020MNRAS.492...72P}. Our estimate for the most compact region, $R \approx 9.8 \times 10^{15}$\,cm, derived from the fastest optical IDV timescale of 0.3\,h shows a remarkable agreement with the size of $R' \approx 5.9 \times 10^{15}$ cm estimated by \citet{2018ApJ...866..102P} using a timing argument on the gamma-ray flare data. This consistency, despite the different wavelengths and variability characteristics (stochastic IDV versus a discrete flare), suggests that simple timing arguments are effective at probing the scale of the most compact, active sites within the jet. In a subsequent work, \citet{2020MNRAS.492...72P} performed detailed SED modeling of the same flare and concluded that a more complex, two-component model was necessary. Their refined model includes a compact inner region with a size of $R' = 1 \times 10^{16}$\,cm and a much larger outer region of $R' = 1.26 \times 10^{17}$\,cm. It is noteworthy that the size of their modeled inner component remains highly consistent with our estimate for the most compact IDV region. This suggests that the rapid optical IDV likely originates from physical regions with dimensions comparable to the compact, primary energy dissipation sites of major gamma-ray flares. The necessity of a second, larger component in the \citet{2020MNRAS.492...72P} model to explain the full SED further strengthens the case for a structured jet, where our IDV analysis effectively isolates the scale of the core emission zone, while a larger volume contributes to the overall flare energetics.
\\
\indent The magnetic field strengths ($B$) derived from the IDV synchrotron cooling timescale arguments in our study range from 0.14\,G to 0.50\,G, with a mean of approximately 0.2\,G. This range is notably consistent with the magnetic field of $B \approx 0.10$\,G derived for Ton~599 by \citet{2018ApJS..235...39C} from their SED fit. However, these values are significantly lower than other estimates derived from broadband SED modeling, such as the $B = 4.0$\,G reported by \citet{2010MNRAS.402..497G} for an average activity state. A particularly relevant comparison can also be made with the studies of the major November 2017 flare. An initial estimate by \citet{2018ApJ...866..102P} based on an equipartition argument yielded a much stronger field of $B_e \approx 6.9$\,G. Subsequent, detailed two-component SED modeling of the same event by \citet{2020MNRAS.492...72P} refined these values to $B=1.3$\,G for an inner component and $B=2.3$\,G for an outer component. While the specific values from different SED models vary, a consistent picture emerges where our IDV-derived magnetic field is weaker than estimates based on the overall energetics of the source. It is important to stress that one-zone models commonly adopted to fit blazar SEDs can be degenerate, which means that multiple choices of the parameters can lead to similarly good results. The differences between the various SED-based estimates (e.g., \citealp{2010MNRAS.402..497G}; \citealp{2020MNRAS.492...72P}) can therefore stem not only from different source activity states or distinct modeling assumptions, but also from this inherent model degeneracy. Despite this, the overarching trend suggests that our IDV analysis consistently probes a weaker magnetic field. This likely reflects the different physical conditions being probed: our IDV-derived $B$ is specific to the rapidly varying, compact subregions, while the global SED-fitting approaches are constrained by the energetics of a larger emission zone. This may imply that our IDV analysis probes localized zones of weaker magnetic fields or enhanced turbulence within an inhomogeneous jet, while the flare and average-state models characterize the stronger, more organized magnetic field of the main jet body.
\\
\indent In addition to constraining the size of the emission region, the rapid variability timescales derived in our work can be used to estimate its distance ($d$) from the central engine. Following the methodology of \cite{2018ApJ...866..102P}, which assumes $d \sim 2c\Gamma^2 t_{\rm var}/(1+z)$, our shortest observed timescale of $\tau_{\rm var} \approx 0.3$\,h (Table 3), combined with a Lorentz factor of $\Gamma=10$ \citep{2017ApJ...846...98J}, places the origin of this fast optical variability at a distance of $d \approx 0.001$\,pc. This finding can be contrasted with the results for major gamma-ray flares in Ton~599 presented by \citet{2018ApJ...866..102P, 2020MNRAS.492...72P}. Their initial analysis, based on a timing argument for a major flare, placed the emission region at $d \approx 0.051$\,pc, a location inside the blazar's BLR, which is estimated to be at $R_{\rm BLR} \approx 0.08$\,pc \citep{2018ApJ...866..102P}. However, in a subsequent, more detailed study involving spectral energy distribution (SED) modeling, the same authors concluded that a simple one-zone model inside the BLR could not self-consistently explain the observed flare properties. Their refined two-component model required emission regions located entirely outside the BLR, at distances of $d \approx 0.097$\,pc and $d \approx 1.22$\,pc, with the external Compton process relying on photons from the dusty torus \citep{2020MNRAS.492...72P}. Our result is therefore highly complementary; it suggests that the most rapid, low-amplitude optical variability we observe originates in a very compact region significantly closer to the central engine than the sites of the more energetic, large-scale gamma-ray flares. This supports a structured jet scenario, where different variability phenomena probe distinct physical locations and scales.

\paragraph{Long-term analysis.} Our extensive 12-year monitoring of Ton 599 (2011–2023) reveals a complex pattern of long-term optical variability, characterized by alternating phases of quiescence and pronounced activity, including several major outbursts (see Figure \ref{fig: F-PD-EVPA_vs_JD} and Figure \ref{fig: STV_outburst}). The fractional variability (F$_{var}$) calculated for the entire period is substantial across all bands, reaching 99\% in the R band for example (Table \ref{tab: LDV_summary}), which is characteristic of highly variable FSRQs (cf. \citealt{2021ApJ...923....7B}, who reports F$_{var}$ values for a sample of $\gamma$-ray bright blazars, with FSRQs such as 3C 279 and 3C 454.3 showing F$_{var}$ of $\sim$80\%).
\\
\indent During periods of enhanced activity, such as the major outbursts in Period II (around JD 2458098 and JD 2458248) and Period III, simultaneous multi-filter observations allowed us to derive an average optical spectral index $\alpha \approx$1.2 (where $F_{\nu} \propto \nu^{-\alpha}$) during the peak of the first major outburst in Period II. This spectral index, corresponding to a relatively flat optical spectrum, is typical for FSRQs during active states when the synchrotron emission from the jet is dominant and can be indicative of fresh particle injection or acceleration. The redshift-corrected monochromatic luminosities derived for these flares, reaching $\log(\nu L_{\nu}) \approx 48.24 \, [\mathrm{erg\,s^{-1}}]$ for the JD 2458098 event, and notably $\log(\nu L_{\nu}) \approx 48.48 \, [\mathrm{erg\,s^{-1}}]$ for the R-band peak of 23.25 mJy during the record-breaking activity in Period III (January 2023), underscore the extreme power of Ton 599. These luminosity values are consistent with those observed in other powerful, high-redshift FSRQs during major flaring episodes (e.g., as is discussed for samples in \cite{2010MNRAS.402..497G}).
\\
\indent The most intense optical activity recorded in our dataset, the R-band peak in January 2023, is particularly noteworthy as it appears to be the optical counterpart to the “brightest ever $\gamma$-ray flare” from Ton 599 observed in January 2023 (MJD 59954.5) reported by \cite{2024MNRAS.529.1356M}. \cite{2024MNRAS.529.1356M} reported a peak $\gamma$-ray flux of $(3.63 \pm 0.20) \times 10^{-6}$ photon cm-2 s-1, and their multiwavelength light curve (their Figure 1) shows simultaneous enhancement in optical/UV and X-ray bands, strongly correlated with the $\gamma$-ray emission. Although a detailed multiwavelength correlation is beyond the scope of this particular paper, the temporal coincidence of our observed record optical flux with this extreme $\gamma$-ray event strongly suggests a common underlying physical origin, likely related to a major energy dissipation event within the jet, such as the propagation of a strong shock or a significant magnetic reconnection event. The complex, often multi-peaked structure of the major outbursts observed in our optical light curves (e.g., Period II and III, see Figure \ref{fig: STV_outburst}) is also a common feature in blazar flares and can be attributed to multiple emission subregions or evolving shock fronts.
\\
\indent Overall, the pronounced long-term optical variations in Ton 599, characterized by significant changes in both flux (spanning over two orders of magnitude) and fractional variability, and culminating in extreme outbursts with luminosities exceeding $10^{48} \mathrm{erg\,s^{-1}}$ , provide critical constraints on jet emission processes. These observations hint at episodic, and at times dramatic, enhancements in the accretion rate or, more likely, in the efficiency of energy transfer to and dissipation within the relativistic jet, driving the observed powerful outbursts across the electromagnetic spectrum.

\paragraph{Color analysis.} The long-term optical color analysis of Ton 599, presented in Sect. \ref{sec: Color analysis}, reveals a remarkably complex spectral evolution that is strongly dependent on the source's brightness level. Our data, spanning 12 years, show three distinct regimes in the CI versus magnitude diagrams, which offer insights into the interplay of different emission components and physical processes.
\\
\indent At faint states, Ton 599 consistently exhibits a RWB trend. This behavior is a well-documented phenomenon in some FSRQs and is typically interpreted as the increasing relative contribution of a relatively constant, redder thermal emission component as the bluer, nonthermal jet emission fades. This thermal component is likely associated with the accretion disk and the BLR. Steward Observatory spectra (Figure \ref{fig: Steward_spectra}) provide direct support for this interpretation, revealing clear thermal emission features, such as the Mg II line and Fe II complexes, particularly during low-flux states. As the source brightens, the nonthermal jet continuum increasingly dominates, diluting these thermal features. A similar RWB trend at faint states, attributed to the thermal emission from the accretion disk (often referred to as the “big blue bump”), was also observed in the powerful FSRQ CTA 102 during its less active phases preceding its major 2016-2017 outburst \citep[see Figure 3 in][]{2017Natur.552..374R}. The study of 3C 454.3 by \cite{2006A&A...453..817V} also implies that at fainter jet states, an underlying, relatively stable “blue bump” component (accretion disk) would be more prominent.
\\
\indent As Ton 599 brightens into intermediate flux levels, its spectral behavior transitions to being largely achromatic, where significant flux variations are not accompanied by systematic color changes. Such achromatic phases in blazars can be more challenging to interpret uniquely but have been attributed to scenarios where the spectral index of the dominant jet emission remains relatively constant during flux changes. Alternatively, geometric effects, such as variations in the viewing angle to different parts of an inhomogeneous or structured jet, or a twisting jet, could lead to flux changes without strong chromatic effects \citep[e.g.,][]{2017Natur.552..374R}. \cite{2017Natur.552..374R} invoked a “twisting inhomogeneous jet” model for CTA 102, where such geometric changes could explain complex variability patterns, including quasi-achromatic periods.
\\
\indent Finally, at its brightest states, Ton 599 displays a clear and statistically significant BWB trend across all color indices. This is the most commonly observed spectral behavior in FSRQs and many BL Lacertae objects during flaring activity \citep[e.g.,][]{2015A&A...573A..69W}. The BWB trend is generally explained by models dominated by synchrotron emission from relativistic electrons within the jet. It can arise from the injection or re-acceleration of electrons to higher energies during outbursts, leading to an increase in the relative contribution of higher-frequency (bluer) synchrotron emission, or a shift of the synchrotron peak frequency toward higher energies. Shock-in-jet models, where shocks accelerate particles and compress the magnetic field, also naturally predict such BWB behavior \citep[e.g.,][]{1998A&A...333..452K,2001ApJ...554....1S,2002PASA...19..138M}. This BWB characteristic was prominently observed in the FSRQ 3C 454.3 during its major optical outbursts, where it was interpreted as the synchrotron jet component, whose peak shifts with flux, overwhelming any constant thermal emission from the accretion disk \citep{2006A&A...453..817V}. The FSRQ CTA 102 also exhibited a strong BWB trend during its exceptional 2016-2017 outburst, consistent with these synchrotron-based models \citep{2017Natur.552..374R}. The overall sequence observed in Ton 599 – from RWB at faint states, through an achromatic phase, to BWB at bright states – strongly supports a two-component model where a relatively stable, redder thermal component is progressively dominated by an increasingly powerful and spectrally hardening (bluer) jet emission as the source brightens. Our estimate of a small underlying thermal disk contribution of 0.13-0.20 mJy is consistent with this picture of jet dominance at higher brightness levels.
\\
\indent Our analysis also revealed significant short-term color variability in Ton 599 on timescales of days to weeks. An intriguing finding is the observed anti-correlation between the amplitude of these short-term CI changes and the average flux density (Figure \ref{fig: dCI_VS_aveFlux}), suggesting that the jet emission is more spectrally stable (i.e., exhibits smaller rapid color fluctuations) during its brightest phases. While detailed modeling of such short-term color evolution is complex, studies of other blazars using high-cadence TESS and ground-based data, such as S5 0716+714  and S4 0954+65, have also revealed intricate and strongly chromatic short-term color variations, typically exhibiting BWB trends \citep[e.g.,][]{2021MNRAS.504.5629R, 2021MNRAS.501.1100R}. More broadly, the blazar literature also documents even more complex phenomena such as “clockwise hysteresis loops” in color-magnitude diagrams during such short-term events. These features are often attributed to the interplay of particle acceleration and cooling timescales within shocks, or the presence of multiple emission components with slight delays.
\\
\indent In summary, the complex and brightness-dependent color evolution of Ton 599 is consistent with a scenario involving contributions from both a thermal accretion disk and BLR component (dominant at faint states) and a powerful, spectrally variable relativistic jet (dominant at bright states). This behavior, including the RWB-achromatic-BWB transitions, mirrors the complexities observed in other well-studied, powerful FSRQs and underscores the dynamic interplay of various emission mechanisms and jet conditions.

\paragraph{Black hole mass.} Our analysis of Steward Observatory optical spectra, obtained during low-flux states of Ton 599 when thermal contributions from the accretion disk and BLR are most discernible, allowed for a robust determination of the central supermassive black hole (SMBH) mass. We carefully measured the Mg II $\lambda$2798 emission line parameters, accounting for complex Fe II blends by fitting the Mg II line with a two-Gaussian model to derive its total flux and FWHM. Using these parameters in the established empirical relation from \cite{2006ChJAA...6..396K}, our estimate for the black hole mass of Ton 599 is log($M_{BH} / M_{\odot}$) = 8.77$\pm$0.26.
\\
\indent This value is consistent with previous estimates for Ton 599 found in the literature, which employ a variety of methods and datasets, though a range of values exists. A summary of these mass estimates is presented in Table \ref{tab:mbh_comparison_no_error}. For instance, \cite{2019PhDT.......127K}, using the Mg II line from DCT spectra, also derived a comparable mass of log($M_{BH} / M_{\odot}$) = 8.95. Broadband SED modeling by \cite{2010MNRAS.402..497G} for their sample of bright Fermi blazars listed log($M_{BH} / M_{\odot}$) = 9.0. Similarly, \cite{2009RAA.....9.1192C}, using SDSS spectra and averaging estimates from H$_{\beta}$ and Mg II lines via empirical relations (\cite{2006ApJ...641..689V} for H$\beta$; \cite{2006ChJAA...6..396K} for Mg II), obtained a value of log($M_{BH} / M_{\odot}$) = 9.11. More recently, \cite{2018ApJS..235...39C}, in a compilation of jet properties for $\gamma$-ray loud AGN, listed a black hole mass of log($M_{BH} / M_{\odot}$) = 8.38 for Ton 599, a value primarily compiled from the literature employing virial methods \cite[e.g][]{2012ApJ...748...49S, 2011ApJS..194...45S}.
\\
\indent Other studies based on single-epoch virial methods have reported somewhat lower, yet broadly consistent, masses. \cite{2006ApJ...637..669L}, employing the H$\beta$ line and an empirical $R_{BLR}-L_{5100}$ relationship, estimated log($M_{BH} / M_{\odot}$) = 8.54. Using HST UV spectra, \cite{2005MNRAS.361..919P} derived log($M_{BH} / M_{\odot}$) = 8.63 from the C IV line and an $R_{BLR}-L_{1350}$ relation. \cite{2005AJ....130.2506X} also reported $log(M_{BH} / M_{\odot})$ = 8.54$\pm$0.21 based on a mass-luminosity relation for L5100 derived from their spectrophotometry of Ton 599. It is noteworthy that \cite{2005AJ....130.2506X} also presented a significantly lower mass estimate of log($M_{BH} / M_{\odot}$) = 7.98 using a method based on the optical variability timescale ($\Delta t_{min}$ > 50 min for Ton 599), which relies on different physical assumptions concerning the innermost stable orbit around a Kerr black hole.
\\
\indent The general consistency of our Mg II-based mass estimate (8.77±0.26) with the majority of other single-epoch virial estimates (typically ranging from log($M_{BH} / M_{\odot}$) $\sim$8.5 to $\sim$9.1) lends confidence to our result. The observed scatter among different literature values is typical for these methods and can be attributed to factors such as the intrinsic variability of the AGN continuum and broad emission lines, different observational epochs and data quality, the specific emission line and calibration used for the $R_{BLR}-L$ relationship, and varying assumptions in SED modeling or alternative estimation techniques. Our derived mass firmly places Ton 599 in the category of hosting a very massive black hole, which is characteristic of luminous FSRQs and radio-loud AGNs, where black hole masses are generally found to be above $10^8 M_{\odot}$. Furthermore, the corresponding Eddington luminosity derived from our mass estimate, $L_{Edd} \approx 7.7 \times 10^{46}$ erg/s, is in line with expectations for such powerful sources, reinforcing the reliability of our spectral measurements and the applied methodology for mass estimation.

\begin{table*}
\centering
\caption{Black hole mass estimates for Ton 599 from the literature and this work.}
\label{tab:mbh_comparison_no_error}
\begin{tabular}{l c l l}
\hline
\hline
Reference & $\log (M_{BH}/M_{\odot})$ & Method & Line(s) Used \\
\hline
This work                                 & 8.77                      & Single-epoch virial (Kong et al. 2006 relation) & Mg II \\
Keck (2017)                           & 8.95            & Single-epoch virial (FWHM, $L_{disk}$) & Mg II \\
Ghisellini et al. (2010) & 9.00                      & SED modeling (disk fitting/estimation) & (Disk) \\
Chen et al. (2009)                           & 9.11                      & Single-epoch virial (avg. of H$\beta$, Mg II) & H$\beta$, Mg II \\
Chen (2018)                   & 8.38                      & Compiled from the literature (virial methods) & (Various) \\
Liu et al. (2006)                            & 8.54                      & Single-epoch virial ($R_{BLR}-L_{5100}$ relation) & H$\beta$ \\
Pian et al. (2005)                           & 8.63            & Single-epoch virial ($R_{BLR}-L_{1350}$ relation) & C IV \\
Xie et al. (2005)                            & 8.54                      & Single-epoch virial ($M-L_{5100}$ relation, corr.) & ($L_{5100}$) \\
Xie et al. (2005)                            & 7.98                      & Optical variability timescale ($\Delta t_{min}$) & (Variability) \\
\hline
\end{tabular}
\end{table*}

\section{Summary and conclusions}

In this work, we have presented a comprehensive analysis of the multiband optical variability of the FSRQ blazar Ton 599, using an extensive dataset collected between 2011 and 2023. We investigated its flux and color evolution across a wide range of timescales, from minutes to years, and used the observations to constrain some of its key physical parameters. Our main findings can be summarized as follows:

\begin{enumerate}
    \item The long-term flux distribution in all observed bands (B, V, R, I) is better described by a log-normal distribution than a normal (Gaussian) one. This finding, supported by a strong linear RMS-flux relation, suggests that multiplicative processes drive the LTV. In contrast, the IDV flux distributions are consistent with a Gaussian profile, pointing to additive processes on the shortest timescales.

    \item The PSD of the light curves follows a red-noise power-law trend, with slopes ranging from $\beta \approx 1.40$ to $1.79$. This confirms that the variability is a correlated stochastic process. We found no statistically significant periodicities in our 12-year dataset.

    \item We detected significant IDV on multiple nights. Using the fastest variability timescales, we constrained the size of the emitting region to the range $R \approx 10^{16}-10^{17}$~cm and the magnetic field strength to $B \approx 0.14-0.50$~G.

    \item Based on the shortest observed optical variability timescale of $\sim$0.3 hours, we estimate that the most rapid, low-amplitude variations originate from a very compact region located at a distance of approximately $d \approx 0.001$~pc from the central engine, placing it significantly closer than the sites of major $\gamma$-ray flares reported in other studies.

    \item Over the monitoring period, the blazar exhibited extreme flaring activity, reaching a peak flux of 23.5~mJy in the R band. This corresponds to an exceptionally high monochromatic luminosity of $\log(\nu L_{\nu}) = 48.48$~[erg~s$^{-1}$].

    \item The blazar's color evolution is complex and strongly depends on its brightness, revealing three distinct regimes: a RWB trend at low fluxes, which we associate with the thermal emission component; an achromatic phase at intermediate brightness; and a dominant BWB trend at the highest flux states, characteristic of the synchrotron jet emission.

    \item Using low-flux optical spectra from the Steward Observatory, we analyzed the Mg~II $\lambda 2798$ emission line. After carefully accounting for Fe~II contamination, we estimated the mass of the central supermassive black hole to be $\log(M/M_{\odot}) = 8.77 \pm 0.26$, a value consistent with previous estimates for this blazar.
\end{enumerate}

Overall, our results underscore the complexity of blazar variability, pointing to a combination of different physical processes acting on different timescales. This comprehensive, long-term photometric and spectroscopic study of Ton~599 provides crucial constraints on the models of particle acceleration and emission in relativistic jets.

\section*{Data availability}
The optical photometric data underlying this article are available in electronic
form at the CDS via anonymous ftp to \url{cdsarc.u-strasbg.fr} (130.79.128.5) or via
\url{http://cdsweb.u-strasbg.fr/cgi-bin/qcat?J/A+A/}.
%Data acquired by the WEBT Collaboration are stored in the WEBT archive and are available upon request to the WEBT President Massimo Villata (\href{mailto:massimo.villata@inaf.it}{massimo.villata@inaf.it}). 

\begin{acknowledgements}

We thank the anonymous referee for constructive comments and suggestions that helped improve the manuscript. C.M.R., M.V. and M.I.C. acknowledge financial support from the INAF Fundamental Research Funding Call 2023. P.K. acknowledges support from the Department of Science and Technology (DST), government of India, through the DST-INSPIRE Faculty grant (DST/INSPIRE/04/2020/002586). We acknowledge the Maidanak Observatory for a sufficient amount of the observations. The research at Boston University was supported in part by the National Science Foundation grant AST-2108622,  and by several NASA Fermi Guest Investigator grants, the latest is 80NSSC23K1507. This study was based in part on observations conducted using the 1.8m Perkins Telescope Observatory (PTO) in Arizona, which is owned and operated by Boston University. This article is partly based on observations made with the IAC80 operated on the island of Tenerife by the Instituto de Astrofisica de Canarias in the Spanish Observatorio del Teide. Many thanks are due to the IAC support astronomers and telescope operators for supporting the observations at the IAC80 telescope. This article is also based partly on data obtained with the STELLA robotic telescopes in Tenerife, an AIP facility jointly operated by AIP and IAC and on observations made with the LCOGT 0.4 m and 1.0 m telescope networks, one of whose nodes is located in the Spanish Observatorio del Teide. O.V. is partially supported by Chinese Academy of Sciences (CAS) President’s International Fellowship Initiative (PIFI) (grant No. 2024VMB0006). G.D., O.V., M.S. M.D.J, J.D.K. and M.L acknowledge support by the Astronomical Station Vidojevica and the Ministry of Science, Technological Development and Innovation of the Republic of Serbia (MSTDIRS) through contract no. 451-03-136/2025-03/200002 made with Astronomical Observatory (Belgrade), by the EC through project BELISSIMA (call FP7-REGPOT-2010-5, No. 256772), the observing and financial grant support from the Institute of Astronomy and Rozhen NAO BAS through the bilateral SANU-BAN joint research project "GAIA astrometry and fast variable astronomical objects", and support by the SANU project F-187. Also, this research was supported by the Science Fund of the Republic of Serbia, grant no. 6775, Urban Observatory of Belgrade - UrbObsBel. L.\v C.P. acknowledge funding provided by the University of Belgrade - Faculty of Mathematics (the contract 451-03-136/2025-03/200104) and Astronomical Observatory Belgrade (the contract 451-03-136/2025-03/200002) through the grants by the Ministry of Science, Technological Development and Innovation of the Republic of Serbia. Observations with the SAO RAS telescopes are supported by the Ministry of Science and Higher Education of the Russian Federation. We acknowledge support by Bulgarian National Science Fund under grant DN18-10/2017 and Bulgarian National Roadmap for Research Infrastructure Project D01-326/04.12.2023 of the  Ministry of Education and Science of the Republic of Bulgaria. This research was partially supported by the Bulgarian National Science Fund of the Ministry of Education and Science under grants KP-06-H68/4 (2022), KP-06-H88/4 (2024) and KP-06-KITAJ/12 (2024). NRIAG team acknowledges financial support from the Egyptian Science, Technology \& Innovation Funding Authority (STDF) under grant number 45779. The Abastumani team acknowledges financial support by the Shota Rustaveli National Science Foundation of Georgia under contract FR-24-515. We acknowledge observations at the Tien Shan Observatory (Kazakhstan). We thank iTelescope for the use of their telescopes that made it possible for us to carry out optical observations. Based on observations made with the Nordic Optical Telescope, owned in collaboration by the University of Turku and Aarhus University, and operated jointly by Aarhus University, the University of Turku and the University of Oslo, representing Denmark, Finland and Norway, the University of Iceland and Stockholm University at the Observatorio del Roque de los Muchachos, La Palma, Spain, of the Instituto de Astrofisica de Canarias. The data presented here were obtained [in part] with ALFOSC, which is provided by the Instituto de Astrofisica de Andalucia (IAA) under a joint agreement with the University of Copenhagen and NOT. James R. Webb is honored to be permitted to conduct astronomical research on Iolkam Du’ag (Kitt Peak), a mountain with particular significance to the Tohono O’odham Nation. Some of the data are based on observations collected at the Observatorio de Sierra Nevada; which is owned and operated by the Instituto de Astrof\'isica de Andaluc\'ia (IAA-CSIC); and at the Centro Astron\'{o}mico Hispano en Andaluc\'ia (CAHA); which is operated jointly by Junta de Andaluc\'{i}a and Consejo Superior de Investigaciones Cient\'{i}ficas (IAA-CSIC). The IAA-CSIC co-authors acknowledge financial support from the Spanish "Ministerio de Ciencia e Innovaci\'{o}n" (MCIN/AEI/ 10.13039/501100011033) through the Center of Excellence Severo Ochoa award for the Instituto de Astrof\'{i}isica de Andaluc\'{i}a-CSIC (CEX2021-001131-S), and through grants PID2019-107847RB-C44 and PID2022-139117NB-C44. Data from the Steward Observatory spectropolarimetric monitoring project were used. This program is supported by Fermi Guest Investigator grants NNX08AW56G, NNX09AU10G, NNX12AO93G, and NNX15AU81G.
\end{acknowledgements}

\bibliographystyle{aa}
\bibliography{ref}

\begin{appendix}

\section{WEBT data}

\begin{figure*}[htbp]
    \centering    
    \includegraphics[width=\textwidth, height=22cm]{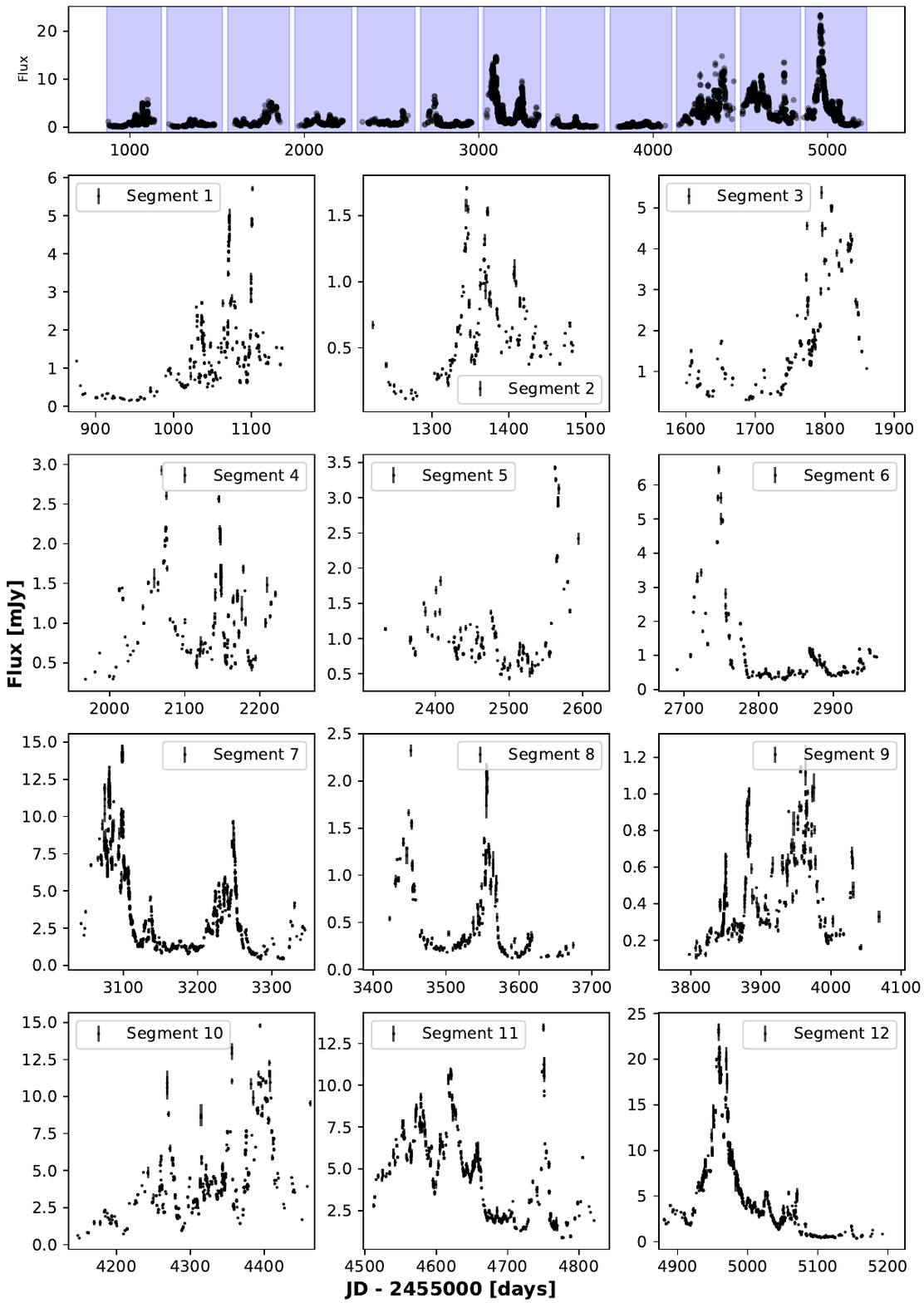}    
    \caption{Light curve segmentation of the blazar Ton 599. 
The top panel shows the full light curve with 12 segments shaded in light blue, numbered from left to right. 
The lower panels provide a detailed view of each segment, arranged in a 4 × 3 grid.}
    \label{fig: plot_outbursts_4x3}
\end{figure*}

The following appendix provides additional information on the data used in this study. Table~\ref{tab: observatories} presents details of the participating observatories and instruments in the WEBT campaign. Figure~\ref{fig: plot_outbursts_4x3} displays the segmentation of the optical light curve into 12 time intervals, which form the basis of the variability analysis discussed in the main text.
\\
\begin{table*}[t]
\tiny
\caption{Optical datasets contributing to this work}
{\centering
\renewcommand{\arraystretch}{1.1} 
\begin{tabular}{llrlc|llrlc}
    \toprule
    \textbf{Observatory} & \textbf{Country} & \textbf{Diameter} & \bm{$N_{obs}$} & \textbf{Symbol} & 
    \textbf{Observatory} & \textbf{Country} & \textbf{Diameter} & \bm{$N_{obs}$} & \textbf{Symbol} \\
    \midrule
    Abastumani           & Georgia    & 70   & 589  & {\color{yellow} $\circ$}  & Perkins              & USA        & 183  & 1436 & {\color{cyan} $\triangle$}    \\
    Belogradchik         & Bulgaria   & 60   & 17   & {\color{blue} $\circ$}    & Roque (JKT)          & Spain      & 100  & 413  & {\color{green} $\triangle$}   \\
    Calar Alto           & Spain      & 220  & 5    & {\color{black} $\circ$}   & Roque (LT)           & Spain      & 200  & 14   & {\color{brown} $\triangle$}   \\
    Cerro Tololo         & Chile      & 100  & 5    & {\color{cyan} $\circ$}    & Roque (NOT)          & Spain      & 256  & 74   & {\color{gray} $\triangle$}    \\
    Cerro Tololo         & Chile      & 40   & 6    & {\color{green} $\circ$}   & Roque (TNG)          & Spain      & 358  & 4    & {\color{magenta} $\triangle$} \\
    Cerro Tololo (SARA)  & Chile      & 60   & 31   & {\color{brown} $\circ$}   & Rozhen               & Bulgaria   & 200  & 36   & {\color{red} $\triangle$}     \\
    Crimean (AP7p)       & Russia     & 70   & 980  & {\color{gray} $\circ$}    & Rozhen               & Bulgaria   & 50/70& 174  & {\color{orange} $\triangle$}  \\
    Crimean (GE)         & Russia     & 70   & 2    & {\color{magenta} $\circ$} & SAO RAS              & Russia     & 100  & 39   & {\color{yellow} +}            \\
    Crimean (ST-7)       & Russia     & 70   & 455  & {\color{red} $\circ$}     & SAO RAS              & Russia     & 50   & 278  & {\color{blue} +}              \\
    Crimean (ST-7;pol)   & Russia     & 70   & 621  & {\color{orange} $\circ$}  & Siena                & Italy      & 30   & 24   & {\color{black} +}             \\
    Haleakala (LCO)      & Havaii     & 40   & 11   & {\color{yellow} $\square$} & Sierra Nevada        & Spain      & 150  & 3    & {\color{cyan} +}              \\
    Hans Haffner         & Germany    & 50   & 675  & {\color{blue} $\square$}  & Sierra Nevada        & Spain      & 90   & 203  & {\color{green} +}             \\
    Kitt Peak (SARA)     & USA        & 90   & 260  & {\color{black} $\square$} & Sirio                & Italy      & 25   & 5    & {\color{brown} +}             \\
    Kottamia             & Egypt      & 188  & 51   & {\color{cyan} $\square$}  & St.Petersburg        & Russia     & 40   & 430  & {\color{gray} +}              \\
    Lowell (DCT)         & USA        & 430  & 43   & {\color{green} $\square$} & Steward (Kitt Peak)  & USA        & 230  & 49   & {\color{magenta} +}           \\
    Lowell (LDT)         & USA        & 430  & 22   & {\color{brown} $\square$} & Steward (Mt. Bigelow)& USA        & 164  & 211  & {\color{magenta} +}           \\
    Lulin (SLT)          & Taiwan     & 40   & 779  & {\color{gray} $\square$}  & Stocker              & USA        & 61   & 104  & {\color{red} +}               \\
    McDonald (LCO)       & USA        & 100  & 4    & {\color{magenta} $\square$} & Teide (IAC80)        & Spain      & 80   & 206  & {\color{orange} +}            \\
    McDonald (LCO)       & USA        & 40   & 15   & {\color{red} $\square$}   & Teide (LCO)          & Spain      & 100  & 5    & {\color{yellow} x}            \\
    MMT                  & USA        & 650  & 3    & {\color{magenta} +}       & Teide (LCO)          & Spain      & 40   & 17   & {\color{blue} x}              \\
    Mt.Maidanak (FLI)\textsuperscript{a}    & Uzbekistan & 150  & 160  & {\color{orange} $\square$} & Teide (STELLA-I)     & Spain      & 120  & 39   & {\color{black} x}             \\
    Mt.Maidanak (Apogee)\textsuperscript{a} & Uzbekistan & 100  & 365  & {\color{yellow} $\triangle$} & Tien Shan            & Kazakhstan & 100  & 49   & {\color{cyan} x}              \\
    Mt.Maidanak (FLI)\textsuperscript{a}    & Uzbekistan & 60   & 1026 & {\color{blue} $\triangle$} & Tijarafe             & Spain      & 40   & 710  & {\color{green} x}             \\
    New Mexico Skies     & USA        & 30   & 19   & {\color{black} $\triangle$} & Vidojevica           & Serbia     & 140/60  & 428/108  & {\color{brown} x}             \\
    \bottomrule
\end{tabular}}
\vspace{0.5em}
\label{tab: observatories}
\tablefoot{Columns list the observatory (with instrument), country, telescope 
diameter (in cm), number of observations ($N_{\mathrm{obs}}$), and the symbol used 
in Figure \ref{fig: F-PD-EVPA_vs_JD}. For the Vidojevica observatory, the entry combines two telescopes 
(140\,cm and 60\,cm) with the corresponding numbers of observations 
(428 and 108, respectively). 
Data from Mt.~Maidanak observatory (marked with $^{a}$) are based on 
\cite{2018NatAs...2..349E}.}
\end{table*}

\end{appendix}

\end{document}